\documentstyle[aps,pre,epsf,eqsecnum]{revtex}

\begin{document}
\twocolumn[\hsize\textwidth\columnwidth\hsize\csname@twocolumnfalse\endcsname
\title{Structural Properties of the Sliding Columnar Phase in Layered
Liquid Crystalline Systems}
\author{L. Golubovi\'{c}$^{1,*}$, T. C. Lubensky$^2$, and C. S. O'Hern$^{2,\dag}$}
\address{$^1$~Department of Physics, Harvard University,
Cambridge, MA 02138}
\address{$^2$~Department of Physics and Astronomy,
University of Pennsylvania, Philadelphia, PA  19104-6396}
\date{\today}
\maketitle

\begin{abstract}
Under appropriate conditions, mixtures of cationic and neutral lipids and
DNA in water condense into complexes in which DNA strands form local $2D$
smectic lattices intercalated between lipid bilayer membranes in a lamellar
stack.  These lamellar DNA-cationic-lipid complexes can in principle
exhibit a variety of equilibrium phases, including a columnar phase in
which parallel DNA strands from a $2D$ lattice, a nematic lamellar phase in
which DNA strands align along a common direction but exhibit no long-range
positional 
order, and a possible new intermediate phase, the sliding columnar (SC) phase,
characterized by a vanishing shear modulus for relative displacement of DNA
lattices but a nonvanishing modulus for compressing these lattices.  We
develop a model capable of describing all phases and transitions among them
and use it to calculate structural properties of the sliding columnar
phase.  We calculate displacement and density correlation functions and
x-ray scattering intensities in this phase and show, in particular, that
density correlations within a layer have an unusual $\exp( - {\rm const.}
\ln^2 r)$ dependence on separation $r$.  We investigate the stability of
the SC phase with respect to shear couplings leading to the columnar phase
and dislocation unbinding leading to the lamellar nematic phase.  For
models with interactions only between nearest neighbor planes, we conclude
that the SC phase is not thermodynamically stable.  Correlation functions
in the nematic lamellar phase, however, exhibit SC behavior over a range of
length scales.\\
{\sl Pacs: 61.30.Cz,61.30.Jf,64.70.Md}
\end{abstract}
\vskip2pc]

\section{Introduction}
\label{sec:intro}
\par
\begin{figure}
\centerline{\epsfbox{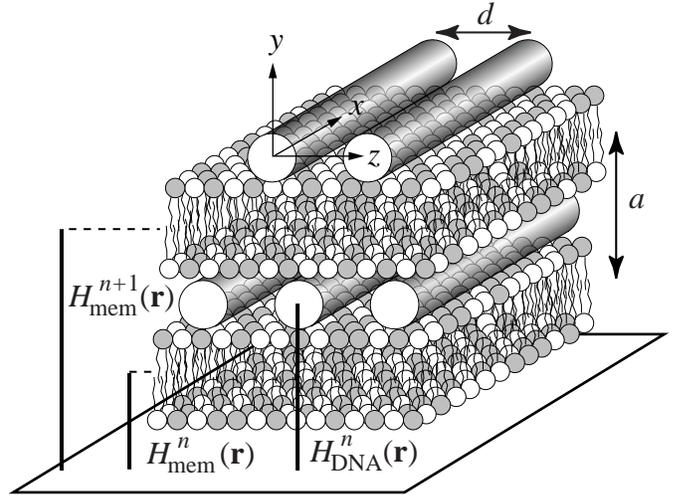}}
\caption{Schematic representation of lamellar DNA-cationic-lipid complexes.
Parallel strands of DNA form $2D$ smectic lattices with lattice spacing
$d$ in galleries between flat lipid bilayer membranes with spacing $a$.  DNA
strands are aligned parallel to the $x$ axis and the $y$ axis is
normal to the lipid planes. The height of the midpoint of the 
$n$th membrane above a
point ${\bf r} = ( x,z)$ in the $xz$-plane is $H_{\rm mem}^n ( {\bf r} )$.  
The $n$th DNA gallery 
with height $H_{\rm DNA}^n ( {\bf r} )$ lies between the $n$th and the $n+1$st
membrane [See. Eqs.\ (\protect\ref{H_mem}) and (\protect\ref{H_DNA})]. }
\label{scfig}
\end{figure}

The search for non-viral vectors for transport of DNA across cell and
nuclear membranes in gene therapy has led to the study of
DNA-cationic-lipid complexes\cite{Felgner}.  DNA and mixtures of
neutral (zwitterionic) and cationic lipids dispersed in water
self-assemble into spheroidal complexes that can attain micron sizes near
the isoelectric point where there is compensation between DNA and lipid
charge.  Fluorescence tagging of DNA and lipids indicate\cite{RadKol97} 
that both species are dispersed uniformly throughout the complexes. 
X-ray scattering experiments reveal two bulk
structures\cite{RadKol97,HTexp,LTexp,SalKol98} for the complexes with close
association between DNA and lipids.  When helper lipids that favor
spontaneous curvature of lipid
membranes or that decrease membrane charge density
are added, a hexagonal inverted micellar structure forms with
DNA molecules captured in the water holes of the hexagonal
lattice\cite{SalKol98}.  When only neutral and cationic lipids are
used, DNA is intercalated in galleries between lamellae of a lamellar
lyotropic phase formed by the lipids\cite{RadKol97,HTexp,LTexp}
as depicted schematically in Fig.\ \ref{scfig}.   This article will
investigate the possible equilibrium phases of these lamellar DNA lipid
complexes.  It will focus primarily on the properties of one phase, the
sliding columnar phase\cite{golubovic_prl,ohern_prl}, 
characterized by strong orientational but weak
positional correlation between DNA strands in neighboring galleries.
\par
DNA molecules are semi-flexible polymers that when confined to a
two-dimensional plane tend to form locally aligned
structures with a preferred inter-molecular separation that can be modeled
as two-dimensional smectic liquid
crystals\cite{2dsmectic,Toner2d,golubovic_pre}.  
Thus, lamellar DNA-lipid complexes can be
viewed as a $3D$ stack of $2D$ smectics as depicted in Fig.\ \ref{scfig}. 
This figure establishes the coordinate conventions that will be used
throughout this paper.  DNA strands
align on average along the $x$ direction, and
the normal to lamellae is along the $y$ axis.  The normal to the $2D$
smectic lattices is along the $z$ direction. The average spacing between
lamellae is $a$, and that between DNA strands is $d$.  The wavenumbers
associated with these lengths are, respectively, $k_0 = 2 \pi/a$  and $q_0
= 2 \pi /d$.   

\begin{figure}
\twocolumn[\hsize\textwidth\columnwidth\hsize\csname@twocolumnfalse\endcsname
\centerline{\epsfbox{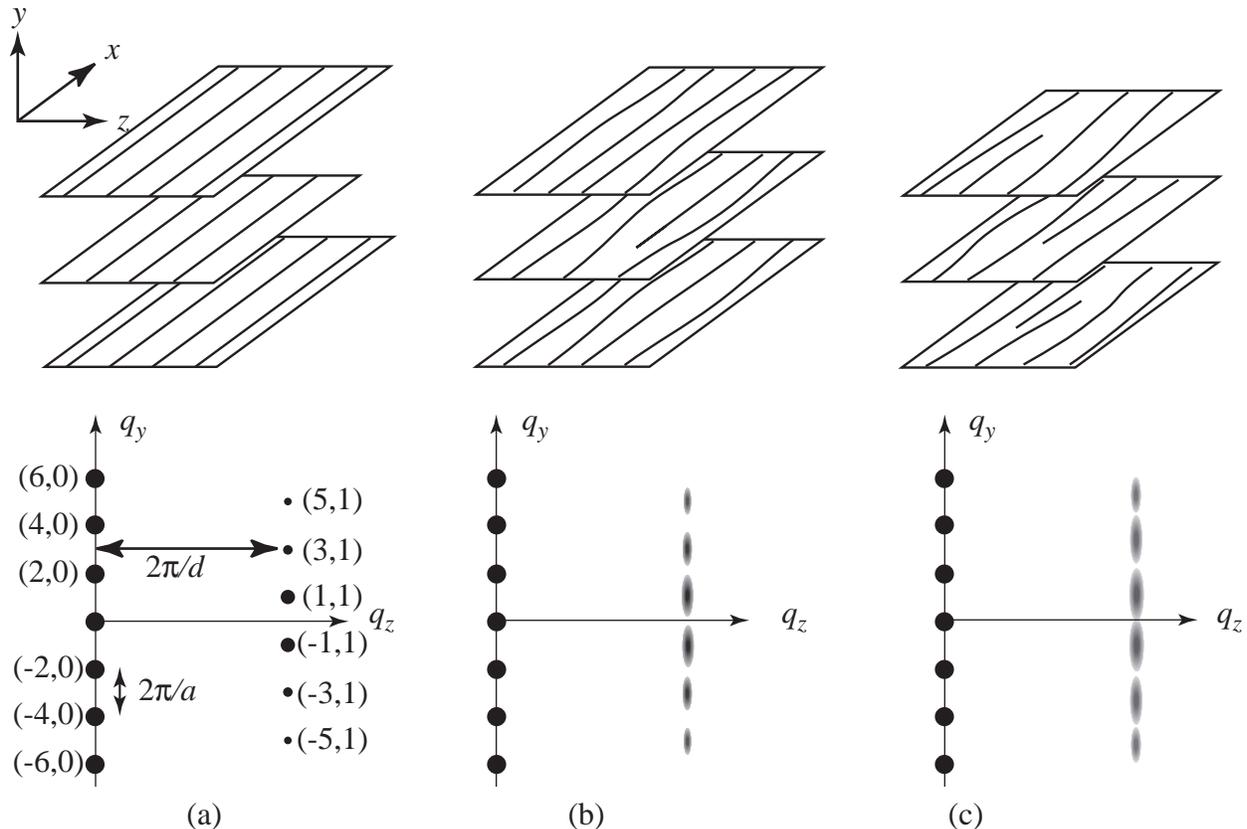}}
\caption{Schematic representation of structure (top) and x-ray scattering
profiles (bottom) of (a) the Columnar phase, (b) the Sliding Columnar
phase, and (c) the Nematic Lamellar Phase.  These figures assume that the
preferred low-temperature phase is the centered-rectangular columnar phase.  
Modifications for a rectangular columnar phase are obvious.
The indexing scheme for the
x-ray intensities is the standard one for centered rectangular lattice.
There are Bragg peaks in the columnar phase at ${\bf G} = (0,m_1 k_0/2, m_2
q_0)$ [denoted in the figure as $(m_1,m_2)$, 
where $m_1 + m_2$ is even].  In the columnar phase,
DNA strands form a $2D$ lattice with true long-range order and
associated x-ray Bragg peaks.  In the sliding columnar and nematic
lamellar phases, there are power-law quasi-Bragg peaks at $(2 m_1,0)$ and
diffuse peaks at $(m_1, m_2)$ for $m_1 \neq 0$.  Dislocations destroy
$2D$ smectic-like order in the nematic lamellar phase to produce more
diffuse $(m_1, m_2)$ peaks than in the sliding columnar phase.
If interlayer couplings are weak and thermal fluctuations sufficiently
strong, the $(\pm m,1)$ peaks may merge to produce a diffuse peak centered
at the rectangular-lattice $(0,1)$ position in the sliding columnar and
nematic lamellar phases. The labelling of peaks in this figure, which follows 
conventional $(x,y,z)$ ordering for the our choice of axes, is the inverse
of that of Ref.\ \protect\cite{LTexp}.  Thus $(m_1,m_2)$ [e.g., $(\pm 1,1)$]
here corresponds to
$(m_2,m_1)$ [e.g., $(1,\pm 1)$] in Ref. \protect\cite{LTexp}.}   
\label{fig:phase-xray}
\vskip2pc]
\end{figure}

If the lamellae
are assumed not to be disrupted by dislocations or rips, then 
the following possible equilibrium phases are easily identified:
\begin{itemize}
\item {\bf Columnar (C) Phase}: In this phase, 
DNA strands are aligned on average
along the $x$ axis, and their centers occupy positions on a $2D$
crystal lattice in the $yz$ plane. Since the standard Coulomb
repulsion between DNA strands favors staggering of smectic lattices in
neighboring galleries, the columnar lattice is normally expected to 
be centered rectangular
as observed in experiments on complexes in which membranes are in the 
$L_{\beta'}$ phase rather than the more disordered $L_{\alpha}$
phase\cite{LTexp}.
We will, however, consider both simple rectangular and centered rectangular
columnar lattices.  A simple rectangular lattice may occur if
the effective interaction between DNA strands in {\it different} galleries
is attractive, as is the case between DNA strands in solutions with
polyvalent salts \cite{DNAattraction}. 
The Columnar phase is favored at low temperature. 
It has the same symmetry as 
a columnar discotic liquid crystal
phase\cite{coldiscotic}. 
It is characterized by a $2D$ elastic energy with a nonvanishing shear
modulus for relative displacements of DNA lattices in different galleries 
and nonvanishing moduli for compression of both lamellae and DNA
smectic lattices.  It has long-range positional order in the
$2D$ $yz$-plane and associated Bragg peaks in its x-ray scattering profile
at reciprocal lattices vectors 
${\bf G}_{m_1,m_2} = [0, m_1 k_0/2, m_2 q_0 ] \equiv (m_1,m_2)$  where $m_1$ and
$m_2$ and integers and $m_1$ is even for a simple rectangular lattice and
$m_1 + m_2$ is even for a centered rectangular lattice 
as shown Fig.\ \ref{fig:phase-xray}.
\item {\bf Nematic Lamellar (NL) Phase}: In this phase, the 
periodic positional order
of the columnar phase is destroyed by dislocations in the DNA smectic
lattices, but the long-range orientational order of DNA strands is
maintained. This phase is characterized by a long-wavelength elastic energy
with a lamellar compression modulus, an anisotropic lamellar bending
modulus, and orientational rigidities (Frank elastic constants) opposing
spatially dependent variation of DNA alignment direction. 
Both the shear modulus for relative displacement of DNA lattices 
in different galleries and the
compression modulus for DNA lattices vanish, 
and there is exponential
decay of DNA positional correlations.  The x-ray scattering profile
of the NL phase exhibits lamellar power-law $(2 l,0)$ peaks at 
${\bf G}_{2l,0} = (0, l k_0, 0 )$.  
If columnar-phase positional correlations are
well developed, it will also exhibit 
Lorentzian peaks at ${\bf G}_{m_1,m_2}$ for $m_2 \neq 0$ as depicted in
Fig.\ \ref{fig:phase-xray}.   
If thermal fluctuations are sufficiently strong that these correlations are
not well developed in centered rectangular systems, then the x-ray
scattering profile could exhibit a broad $(0,1)$ peak in the vicinity of
$(0,0,q_0)$ rather than the expected pair of $(1,1)$ and $(-1,1)$ peaks.  In
this case, the positions of the x-ray scattering peaks would appear to
indicate a tendency to form a simple rectangular rather than the
ground-state centered rectangular structure.
\item {\bf Isotropic Lamellar (IL) Phase}: 
In this phase, orientational as well as
positional order of DNA lattices is lost.  Macroscopically this phase is
identical to a multi-component isotropic lamellar phase.  It is
included for completeness and will not be considered further in this
paper.
\item {\bf Decoupled $2D$ Smectic (DS) Phase}:  
This phase occurs only if there are
absolutely no interactions between DNA lattices in neighboring galleries.
Its elasticity and correlations are thus those of independent $2D$
smectic lattices.  Since there are always interactions between galleries,
this phase will not exist in real systems.  It is, however, a useful
limit to consider since systems with weak coupling between layers will
behave as though they are decoupled at sufficiently short length scales. 
\end{itemize}
In addition to the above phases, which are straightforward to
identify, there is the possibility of another phase with unusual
properties:
\begin{itemize}
\item {\bf Sliding Columnar (SC) Phase}\cite{golubovic_prl,ohern_prl}: 
This phase has properties 
intermediate between those of the columnar and lamellar nematic phases. 
Its elastic energy is distinguished from that of the NL phase by the
presence of a nonvanishing modulus for compression of DNA lattices.
In-plane smectic correlations die off as $\exp (- {\rm const.} \ln^2 r )$
as a function of separation $r$ rather than exponentially as in the NL
phase.  Correlations between smectic lattices in different galleries
die off exponentially with layer-number difference.  The x-ray structure factor 
exhibits power-law lamellar $(0,lk_0,0)$ peaks and well defined DNA
$(m_1,m_2)$ peaks with $m_2\neq0$ that are sharper than the corresponding
Lorentzian peaks in the NL phase, as depicted in Fig.\
\ref{fig:phase-xray}. 
As in the NL phase, $(1,1)$ and $(-1,1)$ peaks may merge to produce a
single $(0,1)$ DNA peak if thermal fluctuations are sufficiently strong.
\end{itemize} 

Rather stringent conditions must be met before the SC phase can be
thermodynamically stable.  Thermal fluctuations must be strong enough to
destroy the interlayer shear coupling present in the columnar phase but not so strong
that dislocation proliferation destroys the smectic compressibility to
create a nematic lamellar phase.  The SC phase is stable only for
temperatures $T$ lying above a decoupling temperature $T_d$ at which the C
phase becomes unstable and below a KT-melting temperature $T_{KT}$ above
which the NL phase becomes stable.  Thus a necessary condition for the SC
phase to be stable is $T_d < T_{KT}$.  In Sec.\ \ref{sec:SCstability}, we
will show that this condition is violated for the nearest neighbor models
we consider here.  Elsewhere\cite{ohernxy,OGLT}, we show that appropriately
chosen interactions between further neighbor planes can stabilize the SC
phase. 

The focus of the paper is the rather unusual properties of the SC
phase.  We will, therefore, assume throughout most of the paper that the SC
phase does exist.  This approach is justified because it can 
under appropriate conditions be an equilibrium phase as discussed in Ref.
\onlinecite{ohernxy} and because even if it is not thermodynamically
stable, there is a range of length scales over which correlations functions
will exhibit SC behavior.

This paper is composed of six sections, of which this is the first, and seven 
appendices, which mostly present mathematical details.  
Section \ref{sec:model} derives the Hamiltonians for the columnar, sliding
columnar, and nematic lamellar phases, based on a model in which the
dominant interactions are couplings between each DNA lattice and the two
membranes on either side of it.  Section \ref{sec:sccorrel} presents the
correlation functions for the SC phase, including x-ray scattering
intensities.  Section \ref{sec:SCstability} addresses the stability of the
SC phase.  It calculates $T_d$, and it derives general expressions for the
interaction between dislocations before calculating $T_{KT}$.  Section
\ref{sec:lengthscales} presents a discussion of the various important
length scales.  Finally, Sec.\ \ref{sec:conclusion} provides an overview
of results.  Appendix \ref{app:Heff} presents details of the derivation of
the Hamiltonians in Sec.\ \ref{sec:model}.  Appendices \ref{app:usquared}
through \ref{app:inpl-corr} provide details of the calculations of various SC
phase correlations functions, and App.\ \ref{app:interaction} presents
details of calculations of interactions between dislocations.

\section{Model for Lamellar Phases of DNA-lipid complexes}
\label{sec:model}
\setcounter{equation}{0}
\subsection{Definition of Variables}
\par
As depicted in Fig.\ \ref{scfig}, lamellar phases of DNA-lipid complexes
consist of a periodic stack of planar lipid bilayer membranes with spacing $a$
separated by galleries intercalated with DNA strands that form a local $2D$
smectic lattice with preferred spacing $d$. Since our primary
interest is in the nature of possible ordering of the DNA strands, we will
assume that the lipid membranes are free of defects such as dislocations or
focal conic structures that destroy their integrity.  We can, therefore,
specify a layer by its integer layer number
$n$, and we can specify positions in the $xz$ plane by the vector ${\bf r}
= (x,z)$.  We take the equilibrium height of the 
midpoint of bilayer membrane $n$ be $na$.  The height of membrane $n$ 
at position ${\bf r}$ is then
\begin{equation}
H_{\rm mem}^n ( {\bf r} ) = na + h^n({\bf r} ) ,
\label{H_mem}
\end{equation}
where $h^n ( {\bf r} )$ is the Lagrangian height variable\cite{EulerLa}
measuring the displacement of layer $n$ from its ideal height.
The $n$th DNA lattice lies between membranes $n$ and $n+1$.  Its
equilibrium height is, therefore, $(n+\case{1}{2})a$, and its height at
position ${\bf r}$ is
\begin{equation}
H^n_{\rm DNA} ( {\bf r} ) = (n + \case{1}{2})a + u_y^n ( {\bf r} ) ,
\label{H_DNA}
\end{equation}
where $u_y^n ( {\bf r})$ measure deviations from equilibrium height.

The DNA 
lattice in layer $n$ can be described in terms of a Fourier expansion of
its density.  It is important to keep track of the relative phases of 
mass-density waves in adjacent layers.  In an ideal columnar lattice, the phase
of the first mass-density wave in layer $n$ is simply $ q_0 (z -
z^n) $, where
\begin{equation}
q_0 = { 2 \pi /d} 
\end{equation}
and $z^n$ specifies the preferred position of lattices in different
galleries.  In a rectangular lattice, $z^n = 0$, and in a centered
rectangular lattice, $z^n = nd/2$.  Phase changes at position ${\bf r}$
relative to the above ideal phases are produced by local translations
described by the Eulerian displacement variable $u_z^n ( {\bf r} )$. 
Thus the DNA density-wave expansion in layer $n$ is 
\begin{equation}
\label{rhoDNA}
\rho_{\rm DNA}^n ( {\bf r} ) = \sum_k \rho^{nk} ( {\bf r} ) e^{i k q_0
[z - z^n - u_z^n ({\bf r} )]} .  
\end{equation}
If each DNA lattice is perfect with identical rods of linear density
$\lambda$ and separation $d$, then $\rho^{nk}( {\bf r} ) = \lambda/d$ for
every $n$ and $k$. Thermal fluctuations will lead to reductions in
$\rho^{nk}$ that
are larger for larger $k$.  Thus, $\rho_{DNA}^n ( {\bf x})$
can be approximated by
\begin{equation}
\rho_{DNA}^n ( {\bf x} ) \approx \rho_0 +
\psi e^{i q_0 [z - z^n - u_z^n ({\bf r})]} + {\rm c.c.} ,
\label{rho1DNA}
\end{equation}
where $\psi$, assumed to be independent of $n$, is the complex 
amplitude of the first non-trivial density wave.
\par
The total mass density of DNA strands at positions
${\bf x} = (x,y,z) \equiv ({\bf r},z)$ is 
\begin{eqnarray}
\rho_{\rm DNA} ( {\bf x} )& = &\sum_{nk}\rho^{nk} ( {\bf r} ) e^{i k q_0
[z - z^n - u_z^n ({\bf r} ) ]} \nonumber \\   
&& \qquad \times f_{DNA}^k [ y - (n+ \case{1}{2})a - u_y^n ({\bf r} )] ,
\label{DNAdensity}
\end{eqnarray}
where $f_{DNA}^k (y)$ is the form factor along the layer normal for a
DNA mass-density wave of wavenumber $kq_0$.  If the DNA stands are
lines with no width, the $f_{DNA}^k ( y ) = \delta(y)$.  If the
strands are modeled as cylinders of radius $r_{DNA}$, then the Fourier
transform of $f_{DNA}^k (y)$ is\cite{HTexp}
\begin{equation}
f_{DNA}^k (q_y)= {2 J_1 (\sqrt{q_y^2 + k^2 q_0^2}~r_{DNA})\over 
\sqrt{q_y^2 + k^2 q_0^2}~r_{DNA}} ,
\end{equation}
where $J_1 (x)$ is the Bessel function of order $1$.
The membrane density is
\begin{equation}
\rho_{\rm mem} ( {\bf x} ) = \sum_n \rho^0_L f_{\rm mem} [ y - na - h^n ( {\bf r}
) ],
\end{equation}
where $\rho_L^0$ is the area density of the lipid and $f_{\rm mem} (y)$ is
the lipid membrane form factor, which typically is equal to
$1/w$ for $-w/2<y<w/2$ where $w$ is the membrane width. X-ray scattering
experiments probe $\langle \rho({\bf q}) \rho ( -
{\bf q} )\rangle$, where $\rho( {\bf q}) $ is the Fourier transform of a linear
combination of $\rho_{\rm DNA} ({\bf x} )$ and $\rho_{\rm mem} ( {\bf x} )$ [See
Sec.\ \ref{x-raysc}].

\subsection{Hamiltonian for Coupled DNA-Lipid Layers}
\par
We are now in a position to derive the Hamiltonian describing elastic
fluctuations of lamellar DNA-lipid complexes.  Our goal is to develop the
simplest model consistent with symmetry.  We begin with individual
membranes and DNA layers.  They are characterized by layer 
bending energies, which can be expressed as
\begin{eqnarray}
{\cal H}^{\rm bend} & = & \case{1}{2}\sum_n a \int d^2 r
\bigl[K_{3d}[(\partial_x^2 + \partial_z^2) h^n ]^2 \nonumber \\
& & \qquad \qquad +K_{2d}[(\partial_x^2 u_y^n )^2 +
(\partial_x^2 u_z^n )^2 ]\bigr] ,
\label{H_bend}
\end{eqnarray}
where $a K_{3d} = \kappa_{\rm mem}$ is the bending rigidity of the 
lipid membranes and $a d K_{2d} = \kappa_{DNA}$ is the bending
rigidity of an individual DNA strand.  We adopt a convention here with a
factor of the layer spacing $a$ multiplying sums over $n$ to
facilitate the continuum limit: $\sum a \int d^2r \rightarrow \int d^3 x$.
Interactions between DNA strands
lead to a preferred separation between strands within a layer 
and to a compressional elastic energy,
characterized by a modulus $B_{2d}$, for changing this separation. 
There is also a preferred
separation between a given DNA layer and the two lipid membranes above and
below it.  Harmonic deviations from this separation are characterized by a
modulus $B_{3d}$. Changing the distance between membranes on either side 
of a DNA layer will lead to an expansion or a 
compression of its DNA smectic lattice, which is described by a coupling with
strength $B_{uh}$ between inplane strain and membrane separation.  
Combining all of
these effects into a single compression energy, we obtain
\begin{eqnarray}
{\cal H}^{\rm com} & = & \sum_n a \int d^2 r 
\bigl[ \case{1}{2}B_{2d} (u_{zz}^n )^2 
\nonumber \\
& & + (B_{uh}/a) u_{zz}^n (h^{n+1} - h^n )
\label{Hcom} \\
& & +(B_{3d}/a^2) 
[(h^{n+1} -u_y^n)^2 + (u_y^n - h^n )^2]\bigr] , \nonumber
\end{eqnarray}
where 
\begin{equation}
u^n_{zz} = \partial_z u^n_z - [(\partial_x u^n_z)^2 
+ (\partial_z u^n_z)^2]/2
\label{2dstrain}
\end{equation} 
is the nonlinear $2D$ Eulerian strain and where factors of $a^{-1}$ were
introduced to facilitate the continuum limit, e.g., $(h^{n+1} - h^n)/a \sim
\partial_y h ( {\bf x} ) $.
There are additional compressional energies involving interactions
proportional to $( h^{n+1} - h^n )^2$, $(u_y^{n+1} - u_y^n )^2$, 
$(u_{zz}^{n+1}-u_{zz}^n )^2$, and other further
neighbor terms.  These are smaller than those considered in Eq.\
(\ref{Hcom}), and we will neglect them.  
They are, however, needed to achieve $T_d < T_{KT}$ and a stable SC
phase\cite{ohernxy}.

Neighboring DNA strands within a gallery prefer to be parallel.  This leads
to a Frank-like orientational energy
\begin{equation}
\label{Horien}
{\cal H}^{\rm orien} = \case{1}{2} \sum_n a \int d^2 r K_z (\partial_z \theta^n
)^2 ,
\end{equation}
where $\theta^n ( {\bf r} ) \approx \partial_x u_z^n ( {\bf r} ) $ is the angle
that the $n$th DNA lattice at ${\bf r}$ makes with the $x$-axis.  In all but
the NL phase, this orientational interaction is subdominant to 
those in Eq.\ (\ref{H_bend}), and we will ignore it.

\begin{figure}
\centerline{\epsfbox{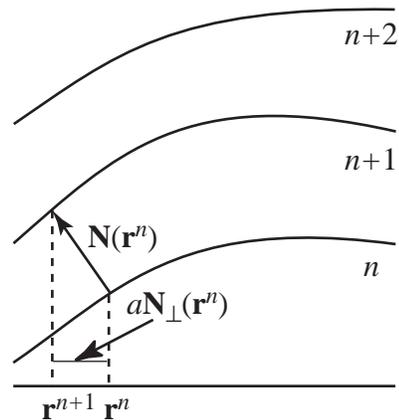}}
\caption{A sequence of tilted and bent membranes. Interactions favor equality 
of the shifted phase, $\beta^n ( {\bf r}) =q_0[ z - u_n ( {\bf r} )]$, in neighboring 
layers at points of closest approach, i.e,, they favor 
$\beta^n ( {\bf r}^n ) = \beta^{n+1} ( {\bf r}^{n+1} )$
where ${\bf r}^{n+1} - {\bf r}^n = a {\bf N}_{\perp} ( {\bf r}^n )$, 
where ${\bf N}_{\perp}$ is the
projection of the membrane layer normal onto the $yz$ plane.  }
\label{fig:nearapp}
\end{figure}

Finally, there are interactions
between DNA lattices in neighboring galleries that favor parallel
alignment and spatial registry of the lattices.  These interactions are
described by a sum of layer-coupling Hamiltonians
\begin{equation}
{\cal H}^{\rm int} = \sum_n ({\cal H}_n^{\theta} + {\cal H}_n^u ) .
\label{H_int}
\end{equation}
The angular coupling is
\begin{equation}
\label{thetacont}
{\cal H}^{\theta}_n = -V^{\theta} \int d^2 r 
\cos[2(\theta^n - \theta^{n+1})] .
\end{equation}
In all but the isotropic
lamellar phase, there is long-range angular order, and we can replace
$\sum_n {\cal H}_n^{\theta}$ by
\begin{equation}
\label{Hrot}
{\cal H}^{\rm rot} 
= \case{1}{2} K_y \sum_n a \int d^2r\left({\partial_x u_z^{n+1}
- \partial_x u_z^n \over a} \right)^2 ,
\end{equation}
where $K_y \approx 4a V^{\theta} e^{-4 \langle (\theta^n)^2 \rangle}$.  
The interaction ${\cal H}_n^u$ in Eq.\ (\ref{H_int}),
favoring spatial registry, arises from interactions between DNA
densities in Eq.\ (\ref{rho1DNA}).  
The phases $q_0 z^n$ [Eq.\ (\ref{rhoDNA})]
are chosen so that energy is minimized when
the remaining phase, $\beta^n ( {\bf r} ) = q_0 [ z - u_z^n ( {\bf r} )]$, of 
mass-density waves in neighboring DNA lattices are equal at the points of
closest approach of the lattices [Fig.\ \ref{fig:nearapp}].  
The point at ${\bf r}^{n+1}$ in layer $n+1$
closest to the point at ${\bf r}^n$ in layer $n$ lies along the normal ${\bf N}^n (
{\bf r} ) \approx ( - \partial_x u_y^n , 1 , - \partial_z u_y^n )$. Thus
${\bf r}^{n+1} = {\bf r}^n + \delta {\bf r}^{n+1}$, where 
$\delta {\bf r}^{n+1} = a {\bf N}_{\perp} \approx ( - a
\partial_x u_y^n, 0 , -a\partial_z u_y^n )$ where ${\bf N}_{\perp}$ is the projection
of ${\bf N}$ onto the $xz$ plane. 
Energy is minimized when
$\beta^{n+1} ( {\bf r}^{n+1} ) = q_0 ( z - a \partial_z u_y^n - u_z^{n+1})$ is
equal to $\beta^n ( {\bf r}^n )$.  The resulting energy is 
\begin{equation}
{\cal H}_n^u = - V^u \int d^2r 
\cos[q_0(u^{n+1}_z - u^n_z + a \partial_z u_y^n)] ,
\label{H_nu}
\end{equation}
where $V^u$ is proportional to the square $|\psi|^2$ of the
mass-density-wave amplitude.  As required, this energy is invariant with
respect to spatially uniform translations described by an $n$- and
${\bf r}$-independent displacement of $u_z^n ( {\bf r} )$.  It is also invariant to
harmonic order with respect to uniform rotations 
in which $u^{n+1}_z - u^n_z = - a \partial_z u_y^n$.  
The predominant interaction between DNA in
neighboring strands is electrostatic, and we expect $V^u \sim (\lambda_e^2
/d) e^{- 2 \pi a /d}$, where $\lambda_e$ is the charge per unit length on a
DNA strand.
\par
To summarize, our complete model Hamiltonian coupling membrane 
height displacements $h^n$
and DNA lattice displacements $u_y^n$ and $u_z^n$ is thus
\begin{equation}
\label{Htot} 
{\cal H}^{\rm tot} = 
{\cal H}^{\rm bend} + {\cal H}^{\rm com} + {\cal H}^{\rm int} ,
\end{equation}
with entries defined by Eqs.\ (\ref{H_bend}), (\ref{Hcom}), and
(\ref{H_int}) to (\ref{H_nu}). 
\subsubsection{Effective Hamiltonian for DNA Displacements}
Though we will discuss membrane height fluctuations, our primary focus will
be on properties of the DNA lattices.  It is, therefore, useful to
integrate out the membrane height variables $h^n ( {\bf r} )$ to obtain an
effective Hamiltonian for DNA displacements only.  This operation is
carried out in Appendix \ref{app:Heff}.  The long wavelength result is
\begin{equation}
\label{Huu}
{\cal H}_{DNA}^{\rm tot} = {\cal H}^{SC} +{\cal H}^u
\end{equation}
where ${\cal H}^u=\sum_n{\cal H}_n^u$ and 
${\cal H}^{SC}$ is the sliding phase Hamiltonian, 
\begin{eqnarray}
\label{HSC}
&&{\cal H}^{SC}  =  \case{1}{2} \sum_n a \int d^2 r [B_{2d} (u_{zz}^n)^2 + K_{2d} (
\partial_x^2 u_z^n ) ^2 ] \nonumber \\
&& + \case{1}{2} \sum_n a\int d^2 r [(B_{3d}/a^2) ( u_y^{n+1} - u_y^n)^2 +
K^{\alpha \beta} (\partial_{\alpha} \partial_{\beta} u_y^n )^2 ]
\nonumber \\
&&+\case{1}{2}\sum_n a\int d^2r[-B_{zz} [(u_{zz}^{n+1} -
u_{zz}^n)/a]^2 \nonumber \\
&&\qquad\qquad+ 2(B_{uh}/a) u_{zz}^n ( u_y^{n+1} - u_y^{n-1} ) ] 
\nonumber \\
&& - \case{1}{2} \sum_n a \int d^2r K_y  [(\partial_x u_z^{n+1} -
\partial_x u_z^n )/a]^2 
\end{eqnarray}
where
\begin{equation}
\label{Bzz}
B_{zz} = {B_{uh}^2\over 4 B_{3d}} ,
\end{equation}
the summation convention on $\alpha = x,z$ and $\beta = x,z$ is
understood, and the bending-rigidity tensor is
\begin{equation}
\label{Kab}
K^{\alpha \beta} = \left(\matrix{K^{xx} & K^{xz} \cr
                                 K^{zx} & K^{zz} \cr}\right) =
                    \left(\matrix{K_{3d} + K_{2d} & K_{3d} \cr
                                 K_{3d} & K_{3d} \cr}\right) .
\end{equation}
Note that integrating out $h^n$ creates interactions between 
nearest-neighbor and next-nearest-neighbor $u_z^n$ and $u_y^n$
displacements not present in the original model of Eq.\ (\ref{Htot}). 
These interactions, however, have particular ratios determined by the
parameters in ${\cal H}^{\rm tot}$.  Of particular importance to us in what
follows is the fixed relations between the coefficients of $(u_y^{n+1} -
u_y^n)^2$, $(u_{zz}^{n+1} - u_{zz}^n )^2$, and $u_{zz}^n (u_y^{n+1} -
u_y^{n-1} )$. In a more general model with further-neighbor interactions,
the simple relation of Eq.\ (\ref{Bzz}) among
$B_{zz}$, $B_{uh}$, and $B_{3d}$ would not hold.   

Using ${\cal H}_{DNA}^{\rm tot}$, we can construct the long-wavelength 
Hamiltonian for each of the phases listed in the introduction as detailed
in the following.

\subsubsection{The Columnar Phase}
The columnar crystal is characterized by strong coupling of displacements
in different layers.  Its elastic Hamiltonian can be obtained by expanding
the cosine in ${\cal H}_n^u$ [Eq.\ (\ref{H_nu})]
about one of its minima.  Performing this
operation, taking the continuum limit, and retaining only the lowest order
terms in a gradient expansion, we obtain the familiar elastic Hamiltonian
for a rectangular columnar lattice\cite{rectco}: 
\begin{eqnarray}
{\cal H}^{C}& = &\case{1}{2} \int d^3x [B_{2d} u_{zz}^2  + B_{3d}
u_{yy}^2  +  B_{uh} u_{zz} u_{yy} + B_{yz} u_{yz}^2 ] \nonumber\\
& & + \int d^3 x [ K_{2d} (\partial_x^2 u_z )^2 + 
(K_{2d} + K_{3d}) (\partial_x^2 u_y )^2 ]
\end{eqnarray}
where ${\bf x} = ({\bf r} , na )$, $u_{zz} ( {\bf x} ) = u_{zz}^{n=y/a} ( {\bf r} )$,
and $B_{yz} \approx V^u q_0^2 a^2$.
Note that the shear elastic modulus $B_{yz}$ arises from ${\cal H}_n^u$ and is
zero when $V^u$ is zero.  The cross compression $B_{uh}$-term arises from the
$u_{zz}^n ( u_y^{n+1} - u_y^n )$ term in ${\cal H}_{DNA}^{\rm tot}$, which was generated by the
interaction of $u_{zz}^n$ with the membrane height field $h^n$

\subsubsection{Sliding Columnar Phase}

The sliding columnar phase is characterized by a vanishing shear modulus
$B_{yz}$ for relative displacement of DNA lattices.  Its dominant
fluctuations are, therefore, described by the Hamiltonian ${\cal H}^{SC}$ of
Eq.\ (\ref{HSC})  obtained by setting $V^u=0$ in ${\cal H}_{DNA}^{\rm tot}$.  The $V^u$
coupling is irrelevant when the sliding phase is stable as detailed in the
following [See Sec.\ \ref{sec:SCstability}]. 

\subsubsection{Nematic Lamellar Phase}
Positional couplings between layers and the smectic compression
modulus $B_{2d}$ both vanish in the nematic lamellar phase, and 
\begin{eqnarray}
{\cal H}^{\rm NL} & = & \case{1}{2} \int d^3 x [ K_x (\partial_x \theta
)^2 + K_y (\partial_y \theta )^2 + K_z (\partial_z \theta )^2 ] \nonumber
\\
& & + \case{1}{2} \int d^3x [ B_{3d} u_{yy}^2 +
+  K^{\alpha\beta} (\partial_{\alpha} \partial_{\alpha} u_y )^2 ] .
\end{eqnarray}
where $K_x= K_{2d}$ and $K_z$ is a Frank elastic constant introduced in
Eq.\ (\ref{Horien}). 

\section{Correlations in the Sliding Columnar Phase}
\label{sec:sccorrel}
\setcounter{equation}{0}

In the preceding section, we derived a general Hamiltonian capable of
describing the phases of lamellar DNA-lipid complexes.  In this section, we
will investigate correlations in the sliding columnar phase.  The most
unusual correlations in the sliding columnar phase are those involving
displacements of DNA strands within the layers.  We will thus begin by
considering the effective Hamiltonian for $u_z^n$ obtained by integrating
out $u_y^n$ from ${\cal H}^{SC}$ [Eq.\ (\ref{HSC})].  From this we will 
calculate all correlation
functions of $u_z^n$.  We will then consider membrane height correlations
described by an effective Hamiltonian for $h^n$
in which $u_y^n$ and $u_z^n$ have
been integrated out of ${\cal H}^{\rm tot}$ [Eq.\ (\ref{Htot})].  
In the long-wavelength limit fluctuations in $u_y^n$ are the same as those
in $h^n$.
Finally, we will discuss the relevancy of the
shear coupling $V^u$, Eq.\ (\ref{H_nu}).  This involves a consideration of both
$u_z^n$ and $\partial_z u_y^n$.
\subsection{Correlations in $u_z^n$}
\subsubsection{Effective Hamiltonian for $u_z^n$}
Since $u_y^n$ and $u_z^n$ are harmonically coupled in ${\cal H}_{DNA}^{\rm tot}$, we can
integrate over $u_y^n$ exactly to obtain an effective Hamiltonian for
$u_z^n $ fluctuations alone.  This calculation is carried out in 
Appendix~\ref{app:Heff}.
In the long-wavelength limit, the resulting Hamiltonian is
\begin{mathletters}
\label{scenergy}
\begin{eqnarray}
\label{scenergya}
{\cal H}_z^{SC} & = & \case{1}{2} 
\sum_n a \int d^2 r \Bigl[B (u^n_{zz})^2 + 
K (\partial_x^2 u^n_z)^2 \\
\label{scenergyb}
& & \qquad\qquad+  {K_y \over a^2} 
[\partial_x(u_z^n - u_z^{n+1})]^2 \Bigr]
\end{eqnarray}
\end{mathletters}
where $K= K_{2d}$ and
\begin{equation}
B = B_{2d} - (B_{uh}^2 /B_{3d} ) .
\end{equation}
Note that there is no term proportional to $(u_{zz}^{n+1} -
u_{zz}^n)^2$ in Hamiltonian ${\cal H}_z^{SC}$ even though there is one in
${\cal H}^{SC}$.  This
is a result of the special relation [Eq.\ (\ref{Bzz})]
among energy coefficients in ${\cal H}^{SC}$.

The Hamiltonian of Eq.\ (\ref{scenergy}) is the sum of two parts: a sum of
elastic energies for independent $2D$ smectics [Eq.\ (\ref{scenergya})] and a
term coupling angles in neighboring layers [Eq.\ (\ref{scenergyb})].  The $2D$
smectic energy is a function of the full nonlinear strain $u_{zz}^n$.
When $K_y=0$ and there is no coupling between layers, nonlinearities in
$u_{zz}^n$ lead to important renormalizations of the long-wavelength
elastic constant of a $2D$ smectic\cite{golubovic_pre}.  When the interlayer
coupling term is present, nonlinearities in $u_{zz}^n$ also lead to
long wavelength renormalization of elastic 
constants in the SC phase\cite{ohern_pre}.
These renormalizations are only logarithmic, however, and we will ignore
them in this article.

We, therefore, consider the harmonic limit of ${\cal H}_z^{SC}$, which is
conveniently expressed in Fourier space.  Introducing
\begin{equation}
\int {d^3 q \over (2 \pi )^2} = \int_{- \pi /a}^{\pi/a} {d q_y \over 2 \pi}
\int_{-\Lambda_x}^{\Lambda_x}{d q_x \over 2 \pi} \int_{-
\Lambda_z}^{\Lambda_z} {d q_z \over 2 \pi}
\end{equation}
where $\Lambda_x$ and $\Lambda_z \sim 2 \pi/d$ are wavenumber cutoffs, and
\begin{equation}
u_z^n ( {\bf r} ) = \int{d^3 q\over (2\pi)^3} e^{i (q_y na + {\bf q}_{\perp} \cdot
{\bf r} ) } u_z( {\bf q} ) ,
\end{equation}
where ${\bf q}_{\perp} = ( q_x, 0 , q_z )$, we obtain
\begin{equation}   
\label{scenergyq}
{\cal H}_z^{SC}=  \case{1}{2}
\int {d^3 q\over (2 \pi )^3}  [B q_z^2 + K
q_x^4 + K_y ( q_y) q_x^2 q_y^2 ] |u_z ( {\bf q} ) |^2 .
\end{equation}
Here
\begin{equation}
K_y( q_y ) = K_y p( q_y a )
\end{equation}
with 
\begin{equation}
p(q_y a ) = 2 {(1 - \cos q_y a)\over (q_ya)^2} ,
\end{equation}
which tends to unity as $q_ya \rightarrow 0$ so that $K(q_y = 0) = K_y$.

Two important length scales can be obtained from Eq.\ (\ref{scenergy}) by
comparing the orientational interaction energy with the $2D$ smectic
compression and bending energies.  The lengthscales
\begin{eqnarray}
\label{3dlengthscales}
x^*={a \over \mu_y}~~{\rm and}~~z^*={a^2 \over \mu_y^2 \lambda}, 
\end{eqnarray} 
with $\mu_y = \sqrt{K_y/K}$ and $\lambda = \sqrt{K/B}$, separate
two-dimensional from three-dimensional behavior.  At lengthscales
within a gallery less than $x^*$ and $z^*$, the $2D$ compression and
bending energies are large compared to the orientational interaction,
and the DNA lattices behave like independent $2D$ smectics. 
On the
other hand, at lengthscales greater than $x^*$ and $z^*$, the
orientational interaction is significant, and $3D$ sliding behavior
occurs.  

Before proceeding to discuss the sliding phase
correlations, we note that the sliding phase
elastic free energy in Eq.\ (\ref{scenergy}) exhibits
a striking {\it local} (gauge) translational invariance of the form
\begin{equation}
u_{z}(x, z, na)  \rightarrow u_{z}(x, z, na) + f(n) ; 
\end{equation}
here the displacement $f(n)$ is an {\it arbitrary} function of $n$
assuming a different value in each gallery.
In other words,  the DNA lattices in different galleries
can continuously slide relative to each other by arbitrary distances 
with no elastic energy costs .
In a standard columnar phase, this continuous
symmetry is spontaneously broken down by 
the shear coupling in Eq.\ (\ref{H_nu}). 
On the other side, in the sliding phase,
this shear coupling is irrelevant at long scales,
and elastic properties reflect the gauge symmetry in 
Eq.\ (\ref{scenergy}).
This symmetry is entirely responsible for the unconventional
fluctuation behavior of the sliding phase we
discuss in the following. In particular,
we find that in the sliding phase, the
fluctuations of relative displacements
$u_{z}^{n+1}({\bf  r}) -u_{z}^{n}({\bf  r})$
of DNA lattices in neighboring galleries
diverge in thermodynamic limit.
As a reflection of the gauge symmetry Eq.\ (\ref{scenergy}),
a DNA lattice in a gallery is essentially
free to flow relative to DNA lattices
in neighboring galleries.
The sliding phase thus exhibits 
zero macroscopic shear modulus\cite{golubovic_prl,ohern_prl}
\subsubsection{Definitions of Correlation Functions}
\label{subsubsec:correlfn}
 
Correlations in $u_z^n$ follow directly form Eq.\ (\ref{scenergyq}):
\begin{eqnarray}
G_{zz} ( {\bf r} , na ) & = &\langle u_z^n ( {\bf r} ) u_z^0 ( 0 ) \rangle \equiv
\langle (u_z^n ( {\bf r} ) )^2 \rangle - g( {\bf r} , n a ) \nonumber \\
& = & \int {d^3 q \over (2 \pi )^3} G_{zz} ( {\bf q} ) 
e^{i({\bf q}_{\perp}\cdot{\bf r} + q_y na)} , 
\end{eqnarray}
where
\begin{equation}
G_{zz}( {\bf q} ) ={T \over B q_z^2 + Kq_x^4 + K_y q_x^2 q_y^2 p(q_y a) } ,
\end{equation}
\begin{equation}
\langle (u_z^n )^2 \rangle = \int {d^3 q \over (2 \pi)^3} G_{zz}( {\bf q} ) ,
\end{equation}
and
\begin{eqnarray}
\label{grna}
g( {\bf r} , na )& = &\case{1}{2} \langle [ u_z^n ( {\bf r} ) - u_z^0 ( {\bf 0 )}
]^2 \rangle \nonumber \\
& = &\int{d^3 q\over (2 \pi )^3} G_{zz}( {\bf q} ) \Bigl[ 1 -
e^{i({\bf q}_{\perp}\cdot {\bf r} + q_y na)} \Bigr] .
\end{eqnarray}
It is useful to decompose $g( {\bf r} , na)$ into three parts:
\begin{equation}
g( {\bf r} , na ) =  g^{(1)}(na) + g^{(2)} ( {\bf r} ) - g^{(3)} ( {\bf r} , na ) ,
\end{equation}
where
\begin{eqnarray}
\label{g1na}
g^{(1)}(na) & = & 
\case{1}{2} \langle [u_z^n ( {\bf 0}) -u_z^0({\bf 0} )]^2 \rangle
\nonumber\\ 
& = & \int {d^3 q \over (2 \pi )^3} G_{zz} ( {\bf q} ) ( 1 - \cos q_y n a )
\\
\label{g2r}
g^{(2)} ( {\bf r} ) & = & 
\case{1}{2} \langle[u_z^0({\bf r}) - u_z^0 ({\bf 0})]^2 \rangle
\nonumber\\
& = &\int {d^3 q \over (2 \pi )^3} G_{zz} ( {\bf q} ) (1 -
e^{i{\bf q}_{\perp} \cdot {\bf r} })  
\\
\label{g3rna}
g^{(3)} ( {\bf r} , na) & = & 
-\langle [u_z^n ({\bf 0})-u_z^0({\bf 0})][u_z^n({\bf r})-u_z^n({\bf
0})]\rangle  \nonumber\\
& = &\int {d^3 q\over (2\pi )^3} G_{zz} ({\bf q} ) 
\nonumber\\ 
& & \qquad \times ( 1 - \cos q_y na )(1 -e^{i {\bf q}_{\perp} \cdot {\bf r} } ) .
\end{eqnarray}
It is clear that $g^{(1)} ( na ) = g(0,na)$ and $g^{(2)} ( {\bf r} ) = g(
{\bf r} , 0 )$. Each of these functions has a characteristic singular
behavior as a function of system size and separation, which we will
summarize below. Finally, the function 
\begin{equation}
\label{g2rn2}
g^{(2)} ( {\bf r} , na ) = g^{(2)} ( {\bf r} ) - g^{(3)} ( {\bf r} , na )
\end{equation}
will appear in our derivation 
in App.\ \ref{app:inpl-corr} of interplane density correlations 
discussed in Sec.\ \ref{x-raysc}.

\subsubsection{Asymptotic Forms for $K_y \neq 0$}

The local fluctuation $\langle (u_z^n)^2 \rangle$ in the SC phase
diverges as the {\em square} of the log of the system size  
with a functional form that depends on the order in which the sample
dimensions $L_x$, $L_y$, and $L_z$ along the $x$, $y$ and $z$ directions
approach infinity:
\begin{equation}
\label{uflucts}
\langle (u_z^n)^2 \rangle = l_u^2
\left\{
	\begin{array}{ll} \ln^2 \left[8 L_x/x^* \right] & L_z
		\gg L_y \sim L_x\\ 
		{1\over 2} \ln^2 \left[\alpha_z L_z/z^*
		\right]& L_x \gg L_y\sim
		L_z  \end{array} \right. ,
\end{equation}
where the length $l_u$ is defined via
\begin{equation}
\label{ludefine}
l_u^2 = {T \over 4 \pi^2 \sqrt{B K_y}},
\end{equation}
$\alpha_z$ is a number, $L_z \gg z^*$, $L_x \gg x^*$, and terms that
do not diverge with system size have been dropped.  The calculation of
the displacement fluctuations in the limit $L_z \rightarrow \infty$
and $L_x \sim L_y$ is detailed in Appendix~\ref{app:usquared}.
Thus, SC ``in-plane" fluctuations are less divergent than $2D$ smectic
fluctuations that scale as a power-law with system size but more divergent than
$3D$ Landau-Peierls smectic lamellar fluctuations that grow logarithmically with
system size\cite{Toner2d,golubovic_pre}.   The mean-square angular fluctuation,
$\langle (\theta^n)^2 \rangle = \langle (\partial_x u_z^n )^2 \rangle$, 
is finite, implying
that the SC phase has three-dimensional long-range orientational order.

The $\ln^2 L_{x,z}$ divergence of $\langle (u_z^n)^2\rangle$ is converted
to a $\ln^2 r$ divergence in the function $g^{(2)} ( {\bf r} )$ [Eq.\ (\ref{g2r})]. 
In Appendix~\ref{app:corrfunction}, we outline the calculation of 
$g^{(2)} ( {\bf r} )$ for large $r$.
The results\cite{golubovic_prl,ohern_prl} are
\begin{equation}
\label{correlationform}
g^{(2)}({\bf r}) = l_u^2
	\left\{ \begin{array}{ll}
	\ln^2\left[8 e^{\gamma} x/x^*\right] + C_x & {\rm if}~z=0\\ 
	{1\over 2}\ln^2\left[32 e^{\gamma} z/z^*\right] + C_z
	& {\rm if}~x=0,\\
	\end{array} \right.  
\end{equation}
where $\gamma \approx 0.577$ is Euler's constant and $C_x$ and $C_z$
are constants that depend on $\Lambda_x$ and $\Lambda_z$ but have
well-defined $\Lambda_{x,z} \rightarrow \infty$ limits.  Note that the
only $\Lambda_{x,z}$ dependence of the correlation function in these
two large distance limits is found in the constants $C_x$ and $C_z$.
$C_x$ and $C_z$ are evaluated in Appendix~\ref{app:corrfunction} in
the continuum limit ($\Lambda_{x,z} \rightarrow \infty$) in which we find $C_x
=0$ and $C_z = \pi^2/8$.  

The function $g^{(1)}(na)$ [Eq.\ (\ref{g1na})] is 
by definition zero when $n=0$.  For all
$n\neq 0$, it diverges logarithmically with system size: 
\begin{equation}
\label{g1def}
g^{(1)}(na)  =  
2 l_u^2 S_n(0) \ln \left[ A_n(\Lambda_x,\Lambda_z) {L_x \over x^*}
\right], 
\end{equation}
where 
\begin{equation}
\label{sndef}
S_n(t) = \int_0^{\pi} du {1-\cos(n u) \over \sqrt{t^2 + u^2 p(u)}},
\end{equation}
\begin{equation}
S_n(0) = \sum_{k=1}^n {2 \over 2 k-1},
\end{equation}
and $A_n(\Lambda_x,\Lambda_z)$ 
calculated in App.\ \ref{app:g1} depends on the layer separation $n$ and the
ultraviolet cutoffs but has a well-defined $\Lambda_{x,z} \rightarrow
\infty$ limit that is calculated in Appendix~\ref{app:g1}.

The function $g^{(3)} ( {\bf r} , na )$ [Eq.\ (\ref{g3rna})] is also zero at $n=0$.
It has the same divergences as a function of separation $r$ that $g^{(1)}(na)$ 
has with system size:
\begin{equation}
\label{g3rnev}
g^{(3)} ( {\bf r} , n ) = 2 l_u^2 S_n ( 0) \cases{\ln (D_n x/x^*) & if $z=0$\cr
                              \ln(E_n a z/z^*) & if
$x=0$.\cr} ,
\end{equation}
where $D_n$ and $E_n$ are numerical constants
that have well defined values in the continuum limit, $\Lambda_x, \Lambda_z
\rightarrow \infty$, as discussed in App.\ \ref{app:g3rna}.

It is useful to
summarize the results of the calculations just presented. When
$n=0$, i.e., for points in the same layer,
$g({\bf r} , 0) = g^{(2)}( {\bf r} )$ grows with $r$ as $\ln^2 r$.  When $n\neq0$, i.e.,
for sites in different layers, $g( {\bf r}, na )$ diverges logarithmically with
system size for all
${\bf r}$ because $g^{(1)} ( na ) $ diverges in this way.  If $g^{(1)} ( na)$
is subtracted from $g( {\bf r} , n)$, then the remaining function, 
$g^{(2)} ( {\bf r} , na )$, grows with $r$ as $\ln^2 (r/r_n)$ with $r_n$ depending on
the coefficient of $\ln r$ in $g^{(3)} ( {\bf r} , na )$ when $n\neq 0$. 

\subsubsection{The limit $K_y=0$: $2D$ Smectic Correlations}

When $K_y = 0$, there are no interactions between planes (when $V^u = 0$),
and the system reduces to a stack of independent $2D$ smectic planes.  Since
experiments\cite{HTexp} are consistent with nearly independent $2D$ smectic
layers, we will review here well established 
results\cite{golubovic_pre} for such systems.

Decoupled smectics are described by the Hamiltonian of Eq.\
(\ref{scenergya}) obtained by setting $K_y = 0$ in ${\cal H}^{SC}$. This
Hamiltonian is a function of the full nonlinear $2D$ strain $u_{zz}^n$. 
Nonlinearities lead to important deviations from harmonic behavior.
Since there is no coupling between layers, $G_{zz} ( {\bf q} )$ is independent
of $q_y$, and we can define a $2D$ correlation function $G_2({\bf q}_{\perp} )
= G_{zz}( {\bf q} ) /a $.

At lengthscales less than the nonlinear lengths
\begin{eqnarray}
\label{nonlinlengths}
l_x = {8\pi  K_2^{3/2} \over T \sqrt{B_2}}~~{\rm and}~~l_z 
= {l^2_x \over \lambda},
\end{eqnarray}
where $K_2 = Ka$ and $B_2 = Ba$ are $2D$ bend and compression moduli and
$\lambda = \sqrt{K_2/B_2}$, nonlinearities are unimportant, and $2D$
smectic fluctuations are described by the linearized elastic Hamiltonian
${\cal H}_z^{SC}$ of Eq.\ (\ref{scenergyq}) with $K_y =
0$\cite{golubovic_pre}.  At lengthscales longer than $l_x$ and $l_z$, the
nonlinear terms in the rotationally invariant strain in 
Eq.\ (\ref{2dstrain}) lead to renormalized bending and
compression moduli $K_2({\bf q}_{\perp})$ and $B_2({\bf q}_{\perp})$ 
that, respectively, diverge and vanish at small wavenumber 
${\bf q}_{\perp}$\cite{golubovic_pre}.  Note that the nonlinear
lengths $l_x$ and $l_z$ decrease with increasing temperature, and thus
nonlinearities become important at high temperatures.  In both the harmonic
and nonlinear regimes, the Fourier transformed displacement correlation
function 
$\langle |u^n_z({\bf q})|^2 \rangle$ in
each gallery were expressed as
\begin{equation}
\label{Gq}
G_2({\bf q}_{\perp}) = {T \over B_2({\bf q}_{\perp}) q_z^2 + 
K_2({\bf q}_{\perp})q_x^4} = 
{T \over B_2} l_z^2 {\widetilde q}_x^{-\gamma} Q({\widetilde q}_z/
{\widetilde q}_x^{\nu}),
\end{equation} 
where ${\widetilde q}_{x,z} = q_{x,z} l_{x,z}$ and 
\begin{equation}
{\widetilde q}_x^{-\gamma} Q( {\widetilde q}_z/{\widetilde q}_x^{\nu} ) \sim
\cases{{\widetilde q}_x^{-\gamma}, & ${\widetilde q}_z = 0$\cr
       {\widetilde q}_z^{-\gamma/\nu} & ${\widetilde q}_x = 0$.\cr}
\end{equation}
The scaling form of the correlation function implies that
$K_2(q_x,q_z=0) \sim q_x^{-4+\gamma}$ and $B_2(q_x=0,q_z) \sim q_z^{-2 +
\gamma/\nu}$.  In the harmonic regime where $q_{x,z} l_{x,z} > 1$,
$K_2({\bf q}_{\perp}) = K_2$ and $B_2({\bf q}_{\perp}) = B_2$ 
are constants, and the
scaling exponents are $\gamma = 4$ and $\nu = 2$. In the anharmonic
regime $q_{x,z} l_{x,z} < 1$, the scaling exponents $\gamma$ and $\nu$
were calculated exactly by mapping the $2D$ smectic model with thermal
fluctuations onto the KPZ model in $1+1$ dimensions
\cite{golubovic_pre}.  The exponents in the anharmonic regime are $\gamma =
7/2$ and $\nu = 3/2$.

The mean-square displacement fluctuations diverge in both regimes with
lengths $L_x$ and $L_z$ of the sample in the $xz$ plane:
\begin{equation}
\label{2dmeansquare}
\langle u_z^2({\bf r}) \rangle = \int {d^2 q_{\perp} \over (2\pi)^2} 
G_2 ({\bf q}_{\perp}) =
\lambda^2 {\widetilde L}_x^{2\alpha} f_u^{(1)}({\widetilde L}_z/
{\widetilde L}_x^{\nu}),
\end{equation}
where $u_z({\bf r}) \equiv u_z^n({\bf r})$, $2\alpha = \gamma - 1 -\nu
= 1$ in both regimes, ${\widetilde L}_{x,z} = L_{x,z}/l_{x,z}$,
$f_u^{(1)}(0) = {\rm const.}$, and $f_u^{(1)}(w) \sim w^{2\alpha/\nu}$
as $w \rightarrow
\infty$.  This implies that the Debye-Waller factor 
$\langle \exp[iq_0 u_z] \rangle^2 = \exp[-q_0^2 \langle u_z^2 \rangle]
= 0$ in the limit of infinite system size, and there is no long-range
positional order at any finite temperature in a $2D$ smectic, even when
there are no dislocations.  

Since the mean-square displacement fluctuations diverge as a power-law
with system size, the displacement correlation function
\begin{eqnarray}
\label{2Dposcorr}
g^{2D}({\bf r})& = &{1\over 2} \langle [u_z^n({\bf r}) - u_z^n(0)]^2 \rangle 
\nonumber \\
&= & 
\lambda^2 |{\widetilde x}|^{2\alpha} f_u^{(2)}(|{\widetilde z}|/|{\widetilde
x}|^{\nu})
\end{eqnarray}
diverges algebraically with in-plane separation $r$.  In
Eq.\ (\ref{2Dposcorr}), ${\widetilde x} = x/l_x$, ${\widetilde z} =
z/l_z$, and the scaling behavior of $f^{(2)}_u(w)$ is similar to that
of $f^{(1)}_u(w)$.  In the harmonic limit,
\begin{equation}
q_0^2 g^{2D} ( {\bf r} ) = \cases{(z /\xi_z)^{1/2}, & if $x= 0$; \cr
                               (x/\xi_x), & is $z=0$ , \cr}
\end{equation}
where 
\begin{equation}
\label{xi_z}
\xi_z = {\lambda B_2^2 d^4 \over 4 \pi^3 T^2}
\end{equation}
is correlation length along $z$ and $\xi_x = \lambda (d/\pi)^2 (B_2 / T) =
2\sqrt{\lambda \xi_z /\pi}$ is a
correlation length along $x$.  These correlation lengths are measured in x-ray
scattering experiments at sufficiently short length scales when layers are decoupled.

Fluctuations in the angle $\theta = \partial_x u_z$
are nondivergent because of an additional factor of $q_x^2$ in the
numerator of (\ref{2dmeansquare}). Finite angular
fluctuations imply that $\langle \cos \theta \rangle = \exp[-\langle
\theta^2 \rangle/2]$ is nonzero, and thus there is long-range
orientational order in $2D$ smectics when there are no 
dislocations\cite{Toner2d}.

\subsection{Correlations in $h^n ( {\bf r} )$ and $u_y^n ( {\bf r} )$.}
In the absence of order in the DNA strands, lamellar DNA lipid complexes are
simply multicomponent isotropic lamellar systems with height fluctuations
identical to those of any isotropic lamellar smectic.  The presence of
orientational order in the DNA lattices introduces anisotropy into the
effective bending moduli of the lipid membranes.  By integrating out
$u_y^n ( {\bf r} )$ and $u_z^n ( {\bf r} )$ from the the complete Hamiltonian 
${\cal H}^{\rm tot}$ of
Eq.\ (\ref{Htot}), we obtain in App.\ \ref{app:Heff}
an effective Hamiltonian for height
fluctuations.  In the long-wavelength limit, this effective Hamiltonian is
\begin{eqnarray}
\label{Hh}
{\cal H}_h^{SC} &= & \case{1}{2}\sum_n a \int d^2 r \Bigl[ ( B_{\rm lam}/a^2)
(h^{n+1} - h^n )^2 \nonumber \\
& & \qquad + K^{\alpha\beta} (\partial_{\alpha}
\partial_{\beta} h^n )^2\Bigr] ,
\end{eqnarray}
where
\begin{equation}
B_{\rm lam} = B_{3d} - (B_{uh}^2/B_{2d}) 
\end{equation}
and $K^{\alpha\beta}$ is the bending-rigidity tensor defined in Eq.\ 
(\ref{Kab}). The model in Eq.\ (\ref{Hh}) is an anisotropic version of
the discrete smectic Hamiltonian often used to describe lamellar
systems\cite{BruinSaf1,Hoylst}.   An effective Hamiltonian for
DNA-lattice height displacements $u_y^n ( {\bf r} )$ can be obtained by
integrating over $h^n( {\bf r} )$ and $u_z^n ( {\bf r})$ in ${\cal H}^{\rm tot}$
of Eq.\ (\ref{Htot}). The resulting Hamiltonian is
identical in the long-wavelength limit to ${\cal H}_h$ in Eq.\ (\ref{Hh}).  Thus
correlations in $u_y( {\bf r})$ are identical in this model in the long-wavelength
limit to those in $h^n ( {\bf r})$.

Fluctuations in $h^n$ are determined by the correlation function
\begin{equation}
G_{hh} ( {\bf q} )  =  {T \over B_{\rm lam} q_y^2 p(q_y a ) +
K^{\alpha\beta} q_{\alpha}^2 q_{\beta}^2} ,
\end{equation}
where $K^{\alpha\beta} q_{\alpha}^2 q_{\beta}^2 = K_{3d} (q_x^2 + q_z^2)^2
+ K_{2d} q_x^4$.
This equation implies that $\langle [h^n ( {\bf r} )]^2\rangle$ diverges
logarithmically with system size as it does in ordinary isotropic lamellar
systems. The coefficient of the system-size logarithm depends on the anisotropy
in $K^{\alpha\beta}$.  Similarly, correlations in $h^n ( {\bf r})$ defined via 
\begin{eqnarray}
\label{gh}
g_h( {\bf r} , na )& = & \case{1}{2} \langle [ h^n ( {\bf r} ) - h^0 ( {\bf
0} )]^2 \rangle \nonumber \\
& = & \case{1}{2} \langle [ u_y^n ( {\bf r} ) - u_y^0 ( {\bf
0} )]^2 \rangle \nonumber \\
& = & \int {d^3 q \over (2 \pi )^3} G_{hh} ( {\bf q} )[1 -  
e^{i({\bf q}_{\perp}\cdot {\bf r} + q_y na)} ] 
\end{eqnarray}
diverge logarithmically with $na$ and ${\bf r}$:
\begin{equation}
\label{g_h}
k_0^2 g_h({\bf r} , na ) 
=\cases{\eta_h \ln ( na \lambda_x \Lambda^2) & if ${\bf r} =0$ \cr
        2\eta_h\ln ({\tilde\Lambda}( \theta )r) & if $n=0$ , \cr}
\end{equation}
where $\lambda_x = \sqrt{B/K^{xx}}$, $k_0 = 2 \pi/a$, and
\begin{equation}
\eta_h = {k_0^2 T \over 8 \pi^2 \sqrt{B_{\rm lam} K^{xx}}} 
I(K^{xz}/K^{xx}, K^{zz}/K^{xx} ) ,
\end{equation}
where
\begin{equation}
I(\mu, \nu ) = {1 \over 2 \pi}
\int_{-\pi}^{\pi} {d\theta \over \sqrt{\cos^4 \theta +  
2 \mu \sin^2 \theta \cos^2 \theta + \nu \sin^4 \theta}} .
\end{equation}
In Eq.\ (\ref{g_h}), $\Lambda$ and ${\tilde \Lambda} ( \theta )$ are cutoffs that
depend on $\Lambda_x$, $\Lambda_z$ and ratios of bending moduli.  In addition,
${\tilde \Lambda} ( \theta) $ depends on the angle $\theta $ that ${\bf r}$ makes with the
$x$ axis. 
\subsection{Density Correlations}
\label{sec:densitycorr}
The DNA density correlation function  arising from displacements parallel
to lipid membranes is
\begin{equation}
S( {\bf r} , na ) 
 = \langle \exp(iq_0 [u_z^n ( {\bf r} ) - u_z^{0} ({\bf 0}) ] \rangle.
\end{equation}
This function is easily evaluated in the SC phase when $V^u = 0$:
\begin{equation}
\label{Srna}
S( {\bf r} , n a ) = e^{- q_0^2 g( {\bf r} , na ) } ,
\end{equation}
where $g( {\bf r} , na )$ is the displacement correlation function defined in
Eq.\ (\ref{grna}).  Since $g( {\bf r} , na)$ diverges with the system size for
all $n\neq 0$, $S( {\bf r} , na)$ vanishes in the thermodymic limit for all
$n\neq 0$.  Thus, DNA densities in different layers are completely
uncorrelated in the SC phase when the coupling $V^u$ in Eq.\ (\ref{H_nu})
is set to zero.  
\subsubsection{In-plane Correlations}
When $n=0$, $g( {\bf r} , na )$ does not diverge in the thermodynamic limit, and we
have
\begin{equation}
\label{S2r}
S_2 ( {\bf r} ) \equiv S({\bf r} , 0)=
\langle e^{i q_0 [u_z^n ( {\bf r} ) - u_z^n ( 0 ) ]} \rangle =
e^{- q_0^2 g^{(2)} ( {\bf r} )} ,
\end{equation}
with $g^{(2)}({\bf r})$ given by Eq.\ (\ref{g2r}).  Thus, from Eqs.\
(\ref{correlationform}) and (\ref{2Dposcorr}), 
\begin{equation}
\label{Sx}
S_2(x,0) =
\cases{S_x e^{-q_0^2 l_u^2 \ln^2[8e^{\gamma} x/x^*]}, & $x \gg x^*$;\cr
e^{-q_0^2 g^{2D}(x,0)}, & $x \ll x^*$,\cr}
\end{equation}
where $S_x = e^{-q_0^2 l_u^2 C_x}$ is a constant, and $g^{2D} ( {\bf r} )$ is
the displacement correlation function of a $2D$ smectic.
In the other direction,
\begin{equation}
\label{Sz}
S_2(0,z) =
\cases{S_z e^{- (q_0^2 l_u^2/2)
\ln^2[32e^{\gamma} z/z^*]}, & $z \gg z^*$;\cr
e^{-q_0^2 g^{2D}(0,z)}, & $z \ll z^*$,\cr}
\end{equation}
where $S_z = e^{-q_0^2 l_u^2 C_z}$ is a constant.  

\subsubsection{Inter-plane Correlations}

As we have just seen $S( {\bf r} ,  na )$ is zero when $n\neq 0$ if $V^u =
0$.  When $V^u$ is not zero, $S( {\bf r} , na) $ is not zero, and it can be
calculated perturbatively in a expansion in $V^u/T$.  Using the
decomposition of Eq.\ (\ref{Huu}) of ${\cal H}_{DNA}^{\rm tot}$ 
in to ${\cal H}^{SC}$ and
${\cal H}^u$, we can express $S( {\bf r} , na)$ as
\begin{equation}
\label{Srn}
S( {\bf r}, na) =  
{ \langle e^{-{\cal H}^u/T} 
e^{iq_0[u_z^n({\bf r}) - u_z^0(0)]} \rangle \over \langle 
e^{-{\cal H}^u/T}  \rangle } ,
\end{equation}
where $\langle \cdot \rangle$ signifies an average with respect to
${\cal H}^{SC}$.  This expression is easily expanded in a power series in
$V^u/T$.  The evaluation of even the lowest order term in this expansion
cannot be carried out analytically.  In App.\ \ref{app:inpl-corr}, we
evaluate the lowest order term in an approximation in which $K_y =0$,
$B_{uh} = 0$, and correlations in $\partial_z u_y^n ( {\bf r} ) $ in different
layers are ignored.  The result for the Fourier transform of $S( {\bf r} ,
na )$ is
\begin{equation}
\label{Sq}
S( {\bf q}_{\perp} , n a ) = \left({{\tilde V}^u\over 2 T}\right)^n
[S_2 ( {\bf q}_{\perp} )]^{n+1} ,
\end{equation}
where ${\tilde V}^u = V^u e^{- W_y}$ with 
\begin{equation}
W_y = q_0^2 a^2 \langle ( \partial_z u_y^n )^2 \rangle /2
\end{equation}
and 
\begin{equation}
\label{S2q}
S_2 ( {\bf q}_{\perp} ) =\int{d^2 q_{\perp} \over (2 \pi)^2} e^{- i {\bf q}_{\perp}
\cdot {\bf r}} S_2 ( {\bf r}) 
\end{equation}
is the Fourier transform of the
two-dimensional intra-plane density correlations function of Eq.\
(\ref{S2r}).
Since the expression, Eq.\ (\ref{Sq}), for $S({\bf q}_{\perp}, na )$ was
derived under the assumption that $K_y = 0$, $S_2 ( {\bf r} ) $ is strictly
speaking the density correlation of an isolated $2D$ smectic.  However,
a more sophisticated variational approximation, to be presented in a
separate publication\cite{OGLT}, yields a result very similar to Eq.\
(\ref{Sq}) for $S({\bf q}_{\perp} , na )$ with $S_2 (
{\bf q}_{\perp})$ replaced by the true intra-plane correlation function with
$K_y \neq 0$ obtained by Fourier transforming $S({\bf r} , 0)$ of Eq.\
(\ref{Srna}).

Transforming $S( {\bf q}_{\perp}, na )$ back to real space, we obtain
\begin{equation}
\label{Srna2}
S( {\bf r} , na ) = \left({{\tilde V}^u \over 2 T}\right)^n \int{d^2
q_{\perp} \over ( 2 \pi )^2 }e^{i {\bf q}_{\perp} \cdot {\bf r}} e^{(n+1) \ln
S_2 ( {\bf q}_{\perp} )} .
\end{equation}
The large $n$ limit of this expression can be calculated using steepest
descents.  For fixed ${\bf r}$ and $n\rightarrow \infty$, the result is 
\begin{equation}
S( {\bf r} , na ) = {2 \pi T \over {\tilde V}^u n \xi_x \xi_z}
e^{-|n| a/\xi_y} e^{-[(x^2 /\xi_x^2 ) + (z^2 /\xi_z^2)]} ,
\end{equation} 
where
\begin{equation}
\label{xixz}
\xi_{\alpha}^2 = {1 \over S_2({\bf r} = 0)}\int d^2 r r_{\alpha}^2 S_2 ( {\bf r} )
\end{equation}
for $\alpha = x,z$, and $r_{\alpha} = x,z$ and
\begin{equation}
\xi_y = a \ln \left({2 T\over {\tilde V}^u S_2 ({\bf q}_{\perp} = 0 )}\right)^{-1} .
\end{equation}
Since $S_2 ( {\bf r} )$ dies off at least as fast as $e^{-r}$ when $K_y
= 0$\cite{Toner2d,golubovic_pre} and as $\exp( -{\rm const.}\ln^2 r)$ when $K_y
\neq 0$, the integrals in Eq.\ (\ref{xixz}) converge, and $\xi_x$ and $\xi_z$
are well defined.

\subsection{X-ray Scattering}
\label{x-raysc}

X-ray scattering experiments probe $I({\bf q} )= \langle \rho ( {\bf q} ) \rho ( -
{\bf q}) \rangle / V$, where $V$ is the volume of the system, and $\rho ( {\bf q} )$
is the weighted sum of the DNA and membrane densities $\rho_{DNA} ( {\bf x} )$
and $\rho_{\rm mem} ({\bf x} )$.  The total scattering intensity can thus be
broken up into a DNA contribution $I_{DNA} ({\bf q})$, a membrane contribution
$I_{\rm mem} ( {\bf q} )$, and a DNA-membrane cross term.  Scattering from the
ordinary columnar phase will exhibit true Bragg peaks at ${\bf G}_{lm} =
(0, (l + \case{1}{2} \sigma m) k_0 , m q_0 )$.  Here $\sigma = 0$ for a simple
rectangular columnar lattice, and $\sigma=1$ for a centered rectangular
columnar lattice.  Since there is no long-range positional order in the SC phase,
there will be no Bragg peaks in its x-ray scattering profile. Rather there will
be membrane-dominated lamellar peaks at 
${\bf G}_{l0} = (0, l k_0, 0 )$
and DNA-dominated sub-power-law peaks at ${\bf G}_{lm}$ for $m\neq 0$.

The contribution of lipid membranes to the x-ray scattering intensity is
\begin{eqnarray}
I_{\rm mem}({\bf q} )& = &{1\over a}|f_{\rm mem} ( q_y )|^2 \nonumber \\
&& \times\sum_ne^{-iq_y na}
\int d^2 r e^{-i {\bf q}_{\perp} \cdot {\bf r}} S_h ( {\bf r}, na, q_y ),
\end{eqnarray}
where 
\begin{eqnarray}
S_h ( r , na , q_y ) & = &\langle e^{-i q_y [h^n ( {\bf r} ) - h^0 ( 0 )
]}\rangle \nonumber \\
& = & e^{- q_y^2 g_h ( {\bf r} , na )} .
\label{S_h1}
\end{eqnarray}
and $g_h( {\bf r} , na )$ is evaluated in Eq.\ (\ref{gh}).
Thus, for example,
\begin{equation}
S_h ( 0, na, k_0 ) \sim (na)^{ - \eta_h} 
\label{S_h2}
\end{equation}
Except for some anisotropies, this
yields power-law Bragg peaks that are essentially identical to those of a
normal lamellar smectic.

The contribution of the DNA lattices to the x-ray scattering intensity is
\begin{eqnarray}
I_{DNA} ( {\bf q} ) & = & \langle \rho_{DNA} ( {\bf q} ) \rho_{DNA} ( - {\bf q} )
\rangle \nonumber \\
& = & {1 \over a} \sum_{nk} |f_y^k(q_y)|^2 |\rho^{nk}|^2 e^{-iq_y na}
e^{- ikq_0 z^n} \nonumber \\
& & \qquad \times \int d^2 r e^{-i \Delta {\bf q}_{\perp}^k \cdot {\bf r}} S_{yz}({\bf r} ,
na , q_y , k) ,
\label{IDNA}
\end{eqnarray}
where $\Delta {\bf q}_{\perp}^k = {\bf q}_{\perp} - k q_0 {\bf e}_z$ and
\begin{equation}
S_{yz} ( {\bf r} , na , q_y , k ) = \langle e^{-iq_y [u_y^n ( {\bf r} ) - u_y^n (
0)]} e^{- i q_0 k [ u_z^n ( {\bf r} ) - u_z^0 ( 0 ) ]} \rangle .
\label{Syz1}
\end{equation}
This correlation function is even more complicated to calculate than $S({\bf r}
, na )$ [Eq.\ (\ref{Srn})].  When $B_{uh} = 0$, $u_y^n$ and $u_z^n$
decouple, and when $k=\pm1$ we have 
\begin{equation}
S_{yz} ( {\bf r} , na , q_y , k=\pm 1) = S_h ( {\bf r} , na , q_y ) S( {\bf r} , na ) ,
\label{Syz2}
\end{equation}
where we used the fact that correlations in $u_y^n$ are the same as those in
$h^n$.
Thus, at ${\bf r} = 0$, this function dies off as 
$n^{- \eta_h -1}e^{- na/\xi_y}$ when $q_y = k_0$.

If we ignore the $S_h ( {\bf r} , na , q_y ) $ contribution of $S_{yz}$
(which would be justified for $B_{\rm lam} \rightarrow \infty$) and we
use the approximation of Eq.\ (\ref{Srna2}) for $S( {\bf r} , na )$, then the
sum over $n$ in Eq.\ (\ref{IDNA}) can be carried out exactly.  The
result for the dominant contribution with $k = \pm 1$ is
\begin{equation}
\label{I_DNA}
I_{DNA} ( {\bf q} ) = { 1 \over a} S_2 ( \Delta {\bf q}_{\perp}^1 ) |f^1_{DNA} (
q_y ) |^2 F( {\bf q} ) ,
\end{equation}
where
\begin{equation}
\label{F(q)}
F({\bf q}) = { 1 - \left( {V^u S_2 (\Delta {\bf q}_{\perp}^1) 
\over 2T}\right)^2 \over 1 - 2 {V^u S_2(\Delta {\bf q}_{\perp}^1) 
\over T} \cos(q_ya - \sigma \pi) + \left({V^u S_2 (\Delta 
{\bf q}_{\perp}^1) \over 2T}\right)^2 }.
\end{equation} 
The structure function $F( {\bf q} )$ reaches a maximum at $q_y = m k_0$ if a
rectangular ($\sigma = 0$) lattice is preferred and at $q_y = (m +
\case{1}{2} ) k_0$ when a centered rectangular lattices ($\sigma = 1$) is
preferred.  At these peaks, $F({\bf q})$ does not exhibit singular behavior:
the function $S_2( \Delta {\bf q}_\perp )$ [Eq.\ (\ref{S2q})] entering Eq.\
(\ref{F(q)}) can be expanded in powers of $\Delta {\bf q}_\perp$ to all orders.

Out-of-plane lamellar fluctuations,
ignored in Eqs.\ (\ref{I_DNA}) and (\ref{F(q)})
(by assuming $B_{\rm lam}\rightarrow \infty$), become
significant when the inter-layer
positional coupling $V^{u}$ is weak.
For example, for $V^{u}=0$ and finite $B_{\rm lam}$,
we find using Eqs.\ (\ref{IDNA}) to (\ref{Syz2}),
\begin{eqnarray}
I_{DNA} &= &{1 \over a} |f^{1}_{DNA}(q_{y})|^2 \nonumber \\
& &\times \int d^{2}re^{-\Delta {\bf q}_{\perp}^{1}\cdot{\bf r}}
S_{h}({\bf r},na=0,q_{y})S_{2}({\bf r}), 
\end{eqnarray}
Using Eq.\ (\ref{S_h1}), it is easy to see that
$I_{DNA}({\bf q})$
has a {\it maximum} at $q_{y}=0, \Delta {\bf
q}_{\perp}^{1}=0$
(even for infinitely thin DNA, with $f^1_DNA(q_y)=1$).
This peak remains at $q_{y}=0$ for
sufficiently weak nonzero $V^{u}$\cite{OGLT}. 
Thus, for sufficiently weak interlayer coupling, 
the scattering
structure factor has a form resembling that of a
simple rectangular
lattice, with a $(0,1)$-like peak at $q_{y}=0$ 
even though the system has
short-range centered rectangular order {\it in real
space}, 
With increasing strength
of inter-layer positional coupling, this peak will
bifurcate, 
at a critical value of $V^{u}$, into two $(\pm 1, 1)$
peaks at nonzero $q_{y}$ [as in Fig.\ \ref{fig:phase-xray}].
Such a change of the form factor is not accompanied by
a thermodynamic phase transition. It is similar in
character to the transition across so-called disordering line
in random microemulsions \cite{micro1,gompschick} at
which the wave-vector maximizing $S({\bf q})$ vanishes.

\section{Stability of the Sliding Columnar Phase}
\label{sec:SCstability}
\setcounter{equation}{0}

Until now, we have treated the sliding columnar phase as though it were
thermodynamically stable.  We will now examine conditions for this
stability.  In order for the SC phase to exist, it must be stable against
forces that bring about registry between neighboring DNA lattices to
produce the columnar phase; and it must be stable with respect to the
disordering effect of dislocations that leads to a nematic lamellar
phase.  In this section, we will investigate these two effects.  We find
that the SC phase orders into the columnar phase via a roughening-like
transition at a decoupling temperature $T_d$ and that it melts to the
nematic phase via a dislocation unbinding transition at a temperature
$T_{KT}$.  Thus, the SC phase can only exist in a temperature range $T_d
< T < T_{KT}$, and it cannot exist at all if $T_{KT} < T_d$.  We
calculate $T_d$ and $T_{KT}$ for our nearest-neighbor model, and we find
that
\begin{equation}
\label{beta}
\beta = {T_{KT}\over T_d} = {1 \over \pi^2} <1 
\end{equation}
for systems with temperature independent coupling constants.
Thus $T_{KT}< T_d$, and, for our model, the SC phase is never
thermodynamically stable.  The introduction of competing nearest-neighbor
and next-nearest-neighbor strain couplings can, however, reverse the
inequality on $T_{KT}$ and $T_d$ and stabilize the SC phase\cite{ohernxy}. 
Even when
the SC phase is not thermodynamically stable, there is a range of length
scales in the lamellar nematic phase in which correlation functions will
exhibit SC behavior.
\subsection{Relevance of Translational Coupling}
\label{reltransc}
In the sliding columnar phase, fluctuations of relative displacements
$u_z^{n+1} ({\bf r}) - u_z^n ( {\bf r})$ of DNA lattices in neighboring
galleries grow with increasing systems size.  We found in Sec.\
\ref{subsubsec:correlfn} that $\langle [u_z^{n+1}({\bf r}) - u_z^n({\bf
r})]^2\rangle \sim \ln L$ [See Eqs.\ (\ref{g1na}) and (\ref{g1def})].   
Due to this divergence, the translational coupling [Eq.\ (\ref{H_nu})] may
become irrelevant above a critical decoupling temperature.
This result is obtained by
calculating the expectation value of the translational coupling
\begin{equation}
\langle {\cal H}^u_n \rangle  = -V^u \int d^2r \langle \cos[q_0 
(u_z^{n+1} - u_z^n  + a \partial_z u_y^n)] \rangle
\end{equation}
with respect to the sliding columnar Hamiltonian in (\ref{scenergy}).  
Because the cross correlation $\langle \partial_z u_y^n ( u_z^{n+1} -
u_z^n)\rangle$ is zero, we can evaluate this quantity exactly: 
\begin{equation}
\langle {\cal H}_n^u \rangle =  -V^u e^{-W_y}  \int d^2 r
\exp[-q_0^2 g^{(1)}(a)] ,
\end{equation}
where $W_y = q_0^2 a^2 \langle (\partial_z u_y^n)^2 \rangle/2$ ,
Using the fact that $g^{(1)}( na )$ diverges logarithmically with system
size [Eq.\ (\ref{g1def})], we find that the translational coupling scales
as 
\begin{equation}
\langle {\cal H}^u_n \rangle \sim -V^u e^{-W_y} L^{2 - \eta_d} ,
\end{equation}
where 
\begin{equation}
\eta_d = 2 (q_0 l_u)^2 S_1(0) = {4 T \over d^2 
\sqrt{B K_y} } .
\end{equation}
The condition $2 - \eta_d = 0$ defines the critical decoupling
temperature 
\begin{equation}
\label{decoupletemp}
T_d = {d^2 \sqrt{B K_y} \over 2}.
\end{equation}  
When $T<T_d$, the translational coupling scales as system size to a
positive power, and $V^u$ is relevant.  In this case, the system becomes a
columnar phase at the longest lengthscales with a nonzero shear
modulus for shifting neighboring lattices relative to each other and
long-range positional order in the $yz$ plane.  When $T>T_d$, the
translational coupling scales as the system size to a negative power,
and $V^u$ is irrelevant. The system flows to the sliding columnar phase at
the longest lengthscales when $T>T_d$. Thus, $T_d$ marks the transition
from the columnar phase to the SC phase as temperature is increased. This
transition is of the roughening type. There is no energy cost for shifting
neighboring lattices relative to each other, and thus the sliding columnar phase
is positionally disordered in the
$yz$ plane.  These conclusions are supported by an RG analysis to be
presented in Ref.\ \onlinecite{OGLT}.

\subsection{Dislocation unbinding}
We have just seen how thermal fluctuations weaken interlayer couplings
to produce the SC phase from the columnar phase. We will now consider
the disordering effects of dislocations.  An individual edge
dislocation in a $2D$ smectic has a finite rather than a logarithmically
divergent energy. As a result, there are thermally excited unbound vortex
pairs at all temperatures that convert the $2D$ smectic into a $2D$ nematic
at the longest lengthscales\cite{Toner2d}.  
In a sliding columnar phase, each DNA
smectic layer experiences an orientational ordering field from its
neighbors.  As a result, the energy of an individual edge dislocation in
a given layer diverges logarithmically with system size, and there can be
a Kosterlitz-Thouless dislocation unbinding transition.  The
low-temperature phase with bound dislocations is the sliding columnar
phase, and the high-temperature phase with unbound dislocations is the
lamellar nematic phase.

\subsubsection{Dislocation Energy}
\label{sec:energy}

The DNA smectic lattice in each layer can have edge dislocation defects
in which the displacement field $u_z^n$ undergoes a change of 
$k d$, where $k$ is an integer,
in one circuit around the defect core.  If there are 
dislocations with integer strengths $k_{n,l}$ at positions
${\bf r}_{n,l}$ in layer $n$, then
\begin{equation}
\label{densdef1}
\bbox{\nabla}_{\perp} \times {\bf v}^n({\bf r}) = 
b^n({\bf r}) {\hat y},
\end{equation}
where ${\bf v}^n({\bf r}) =
\bbox{\nabla}_{\perp} u_z^n({\bf r})$ and 
\begin{equation}
\label{densdef2}
b^n({\bf r}) = d \sum_l k_{n,l} \delta^2({\bf r} - {\bf r}_{n,l})
\end{equation}
is the dislocation density in layer $n$.  

We can now calculate the energy cost of an arbitrary assembly of edge
dislocations in the sliding columnar phase using the SC Hamiltonian
\begin{eqnarray}
\label{gauge}
{\cal H}^{\rm sc} & = & 
{1 \over 2} \sum_n a \int d^2 r \Big[ B(v_z^n)^2
+ K (\partial_x v_x^n)^2 \nonumber \\
&  &\qquad + {K_y \over a^2}(v_x^n - v^{n+1}_x)^2 \Big]
\end{eqnarray}
written in terms of ${\bf v}^n({\bf r})=(v_x^n, v_z^n)$.  
Our strategy is to minimize
this Hamiltonian subject to a nonzero dislocation density $b^n({\bf
r})$.  As usual, the calculation is simpler in Fourier space.  To
transform from real space to Fourier space, we use
\begin{eqnarray}
\label{vqdef}
{\bf v}({\bf q}) & = & \sum_n a \int d^2r e^{-i({\bf q}_{\perp} \cdot {\bf r} 
+ q_yna)} {\bf v}^n({\bf r}), \\
\label{bqdef}
b({\bf q}) & = & \sum_n a \int d^2r e^{-i({\bf q}_{\perp} \cdot {\bf r} 
+ q_yna)}  b^n({\bf r}) \nonumber\\
& = & a d \sum_{n,l} k_{n,l} e^{-i ({\bf q}_{\perp} \cdot 
{\bf r}_{n,l} + q_y na)}. 
\end{eqnarray}
We minimize Eq.\ (\ref{gauge}) using the Euler-Lagrange equations and find  
\begin{eqnarray}
\label{euler}
v_x({\bf q}) & = & - {q_z \over q_x} {B \over  
K({\bf q})q^2} v_z({\bf q}),
\end{eqnarray}
where 
\begin{equation}
K({\bf q})q^2 = K q_x^2 + K_y q_y^2 p(q_y a) .
\end{equation}
We then employ the constraint, Eq.\ (\ref{densdef1}), to relate
$v_z({\bf q})$ to the specified dislocation density $b({\bf q})$.  In
the final step, we insert the expressions for ${\bf v}({\bf q})$ into
the Fourier transformed version of Eq.\ (\ref{gauge}) and find
\begin{equation}
\label{disenergy}
E_D = {B \over 2} \int^{\pi/a}_{-\pi/a} {d q_y \over 2\pi} 
\int {d^2q_{\perp} \over (2\pi)^2} { K({\bf q})q^2|b({\bf q})|^2
\over Bq_z^2 + K({\bf q})q^2 q_x^2 }
\end{equation}
for the energy of edge dislocations in the sliding columnar phase.
Also note that if we set $K_y=0$,
Eq.\ (\ref{disenergy}) reduces to the expression for the energy cost for edge
dislocations in a 2D smectic\cite{Toner2d}.

We now decompose the dislocation energy into a part that
diverges with system size and a part that diverges only with the
separation between defects.  After we insert Eq.\ (\ref{bqdef}) into
Eq.\ (\ref{disenergy}), we find
\begin{eqnarray}
\label{partition}
E_D & = & { \sqrt{B K_y} d^2 \over 4 \pi^2} \Big[
\sum_{n,n'}\sigma_n \sigma_{n'}
J_{n-n'} \ln \left[B_{n-n'}(\Lambda_x) L_x/x^*\right] \nonumber \\
&+ & \sum_{n,n',l,l'} k_{n,l}k_{n',l'} E_{n-n'}({\bf r}_{n,l}-{\bf r}_{n',l'})
\Big], 
\end{eqnarray}
where $\sigma_n = \sum_l k_{n,l}$ is the total dislocation charge
in the $n$th layer, 
\begin{equation}
\label{jndef}
J_n = \int_0^{\pi} du \sqrt{u^2 p(u)} \cos nu = {4 \over 1 - 4n^2}, 
\end{equation}
and $B_n(\Lambda_x)$ is given in Appendix~\ref{app:interaction}.  The
first term diverges logarithmically with system size.  The interaction
energy $E_n({\bf r})$ [Eq.\ (\ref{partition})], 
on the other hand, is written in terms of an
integral that vanishes when ${\bf r} = 0$, and thus it
diverges only with the separation between defects.  In the limit of
infinite system size, $J_{n-n'}$ is a positive definite matrix.  
The dislocation energy diverges logarithmically with system size unless
the total dislocation charge in each layer is zero, i.e.,
$\sigma_n = 0$ for every $n$.

The dislocation interaction energy of Eq.\ (\ref{partition}),
\begin{eqnarray}
E_n({\bf r}) & = & - \int_0^{\pi} du \int_0^{\Lambda_x x^*} {dt \over t}
\sqrt{t^2 + u^2 p(u)} \cos nu \\
& & \times \left(1 - \cos[t x/x^*] \exp\left[
-{z \over z^*} t\sqrt{t^2 + u^2 p(u)}\right] \right) , \nonumber
\end{eqnarray}
is difficult to calculate for arbitrary separations 
$({\bf r},na)$, however, it can be calculated in the limits $x\gg x^*$, $z=0$ 
and $z \gg z^*$, $x=0$.  We find that $E_n({\bf r})$ scales as 
\begin{eqnarray}
\label{logdisenergy}
E_n({\bf r}) = -J_n \left\{ \begin{array}{ll}
\ln [C^x_n(\Lambda_x) |x|/ x^* ] 
&{\rm if}~x\gg x^*~{\rm and}~z=0\\ 
\ln [C^z_n(\Lambda_x) |z|/z^* ]  
& {\rm if}~z\gg z^*~{\rm and}~x=0,\\
	\end{array} \right. \nonumber 
\end{eqnarray}
where $C^{x,z}_n(\Lambda_x)$ are numbers that depend on the layer
index $n$ and the cutoff $\Lambda_x$ and are given in
Appendix~\ref{app:interaction}.  Note that the $\Lambda_x \rightarrow
\infty$ limit of $E_n({\bf r})$ is not well-defined.  Since 
$E_n(r) \sim \ln r/(4n^2-1)$ for large $r$, $E_n({\bf r})$ is positive
for all $n>0$ and negative for $n=0$.  As a result, like-signed
dislocations in different layers {\sl attract} each other, whereas
like-signed dislocations in the same layer {\sl repel} each other.

\begin{figure}
\centerline{\epsfbox{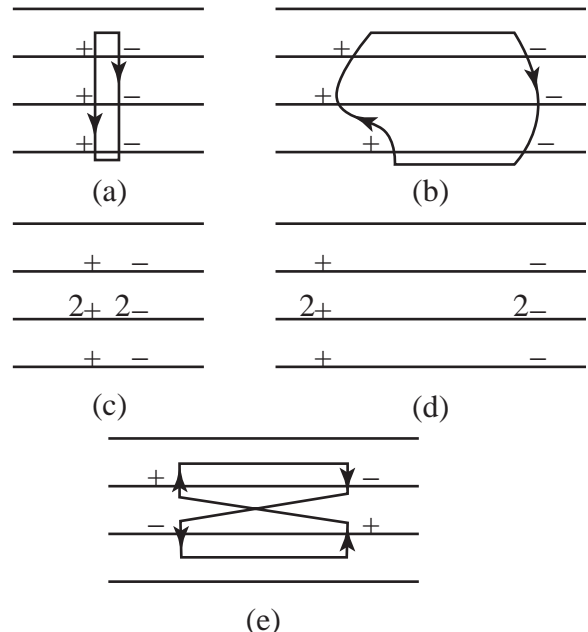}}
\caption{(a) A sequence of closely bound dislocations in neighboring layers
interpreted as a dislocation loop. (b) An unbound version of the
dislocation configuration of (a), also interpreted as a dislocation loop.
(c) a complex composite-dislocation pair. (d) Melted composite pair of (c).
(e) Opposite sign dislocations at nearby points in neighboring layers
interpreted as a crossed dislocation loop.}
\label{fig:disl}
\end{figure}

The attraction of like-sign dislocations in different galleries makes
physical sense.  The dislocation excitations of the SC phase can be viewed
as closed loops carrying a single value of charge
with portions of the loop passing normal to layers and portions
passing parallel to layers.  Those parts of the loops passing through
layers give rise to layer dislocations, and those parts aligned parallel to
layers cost no energy because the shear modulus is 
zero\cite{golubovic_prl,ohern_prl}.  A direction can be assigned to a
loop.  A loop section penetrating a layer in the upward direction gives
rise to a dislocation of one sign while one penetrating in the downward
direction gives rise to a dislocation of the opposite sign. As is apparent
from Fig.\ \ref{fig:disl}, the expectation is that the lowest energy
configuration of a continuous loop will
penetrate nearby points in nearby layers in the same direction, i.e., that   
nearby dislocations in nearby layers will have the same sign.  This
argument is, of course, not rigorous because it is possible for dislocation
lines running parallel to layers to cross to produce nearby dislocations in
nearby layers of opposite sign as shown in Fig.\ \ref{fig:disl}(e).   In
fact, it is possible to construct interlayer interactions for which
like sign dislocations in neighboring layers repel rather than attract.

\subsubsection{Dislocation Unbinding Temperature}
\label{sec:kttransition}

In the previous section, we showed that the dislocation energy is
logarithmically divergent unless there is dislocation charge
neutrality in each layer.  Consider now a configuration in which layer $n$
has a single dislocation of charge $\sigma_n$ (which may be zero).    
The free energy of such a composite dislocation configuration is, by Eq.\
(\ref{partition}),
\begin{eqnarray}
\label{F_D}
F_D &=& E_D - TS_D \\
&=&
\left({ \sqrt{B K_y} d^2 \over 4 \pi^2 T} \sum_{n,n'}
J_{n-n'} \sigma_n \sigma_{n'} - 2T\right) \ln L + {\rm const.} .
\nonumber
\end{eqnarray}
Clearly Eq.\ (\ref{F_D}) indicates that composite dislocations pairs unbind
for temperatures above the Kosterlitz-Thouless type temperature,
\begin{equation}
\label{tkt}
T_{KT}[\sigma_n] = { \sqrt{B K_y} d^2 \over 8 \pi^2} \sum_{n,n'}
J_{n-n'} \sigma_n \sigma_{n'}. 
\end{equation}
If $\sigma_n$ is nonzero only in layer $m$ ($\sigma_n = \delta_{nm}$), the
transition corresponds to unbinding of simple dislocations pairs within a
single layer.  If $\sigma_n$ is nonzero in more than one layer, the
transition corresponds to the unbinding of composite multilayer dislocations
pairs as depicted in Fig.\ \ref{fig:disl}b.  

We see that each configuration of dislocations $\{\sigma_n\}$ will
have a different unbinding temperature.  
The naive expectation is that $T_{KT}$ is lowest when there is a single
dislocation in a single layer.  It is, however, possible that composite
dislocations might melt at a lower temperature.
Hence, we define $T_{KT} =
\min_{\sigma_n} T_{KT}[\sigma_n]$.  Above this temperature, one of the
dislocation configurations (either composite or individual) unbinds,
renormalizes the compression modulus $B$ to zero, and destroys the
in-plane $2D$ smectic order of the sliding columnar phase.  The lowest
unbinding temperature corresponds to the configuration with the lowest
$\ln L_x$ divergent energy [See Eqs.\ (\ref{tkt}) and (\ref{partition})].  
Since $J_n < 0$ for all $n>0$, $|J_n|$
decays with increasing layer separation, and the dislocation energy
scales quadratically with the dislocation charge, the lowest energy
configuration with $N$ defects is a string of $N$ defects of
strength $+1$ with
one in each layer, i.e. $\sigma_n =
\sum_{p=1}^N \delta_{n,p}$.  It is then straightforward to show that
the lowest of these configurations is an individual defect with $N=1$.
Therefore, the dislocation unbinding temperature for the sliding 
columnar phase is 
\begin{equation}
\label{kttemp}
T_{KT} = {d^2 \sqrt{B K_y} \over 2\pi^2}.
\end{equation}
The ratio of this temperature to the unbinding temperature
$T_d$ [Eq.\ (\ref{decoupletemp})] yields 
Eq.\ (\ref{beta}) for $\beta = T_{KT}/T_d = 1 / \pi^2$.

\subsection{Comments on Lyotropic Systems}

An interesting aspect of the present study is
the transitions from the sliding to the columnar
and the nematic phase we discussed above
in terms of the corresponding critical temperatures: the decoupling
temperature $T_{d}$ [Eq.\ (\ref{decoupletemp}], for the columnar to
sliding phase transition, and $T_{KT}$ [Eq.\ (\ref{kttemp})]
for the sliding to nematic phase transition.
In lyotropic systems of interest here [Refs.\
\cite{RadKol97}-\cite{SalKol98},
temperature changes can hardly 
produce significant effects. Nonetheless,
the above transitions can be still triggered by
varying the period of DNA lattices
(as lattices described in Refs.\ \cite{RadKol97}-\cite{SalKol98}).
Indeed, by previous discussions, the decoupling condition
$T>T_{d}$, is equivalent to $\eta_{d}>2$,
with $\eta_{d}=4T/d^{2}(BK_{y})^{ 1 / 2 }$. 
Thus, the decoupling
transition from the columnar to sliding phase
may be triggered 
by changing the strength of the inter-layer orientational
coupling constant $K_{y}$. As evidenced clearly
in experiments \cite{LTexp,HTexp}, the strength of
inter-layer couplings significantly increases  by 
increasing the period of DNA lattices. 
Thus, by the swelling these lattices,
one may reach the transition in which the sliding
phase changes into a columnar phase, as discussed
in Sec.\ \ref{reltransc}, and, in more detail, in Ref.\ \cite{OGLT}.
Finally, we emphasize that, within the simple model
discussed here, there is no true sliding phase, as
suggested by Eq.\ (\ref{beta}) that applies to
systems with temperature independent
coupling constants.
It should be stressed however that the restriction of
having temperature-independent constants
is not essential for this important conclusion 
about the sliding phase stability.
Neither it is essential whether the system is
thermotropic or lyotropic in its character.
Indeed, by Sec.\ \ref{reltransc}, the condition to have dislocations bound 
[i.e., $T<T_{KT}$] is equivalent to $\eta_{KT}>2$,
with $\eta_{KT}=d^{2}(BK_{y})^{1 / 2}/ \pi^{2} T$, 
for the model discussed here [Sec.\ \ref{sec:model}]. Note that
\begin{equation}
\eta_{d}~\eta_{KT}= 4/ \pi^{2} < 4 .
\end{equation}
This relation implies that
the condition for dislocations confinement
($\eta_{KT}>2$) and the condition 
($\eta_{d}>2$) for sliding phase decoupling cannot be simultaneously realized.
Whenever the decoupling condition is realized, 
$\eta_{d}>2$, 
$\eta_{KT}=4/ \pi^2 \eta_{d}<2$, and dislocations will unbind.
This feature turns the sliding phase into a nematic phase
at long length scales. Still, at low dislocation
densities (what may well be the realistic case, see Sec.\
\ref{sec:lengthscales}),
in such a  nematic phase, there is a broad range
of length scales exhibiting sliding-phase correlations
and other structural properties discussed in this work.

\section{Characteristic Lengthscales}
\label{sec:lengthscales}
\setcounter{equation}{0}

Whether or not sliding columnar behavior is seen in recent X-ray
scattering experiments on CL-DNA complexes is still an open
question\cite{HTexp,LTexp}.  Even though the sliding
columnar phase is converted into a nematic lamellar phase at the
longest lengthscales, the SC phase may exist at shorter lengthscales
determined by the density of edge dislocations.  Thus, CL-DNA
complexes studied in recent experiments with domain sizes $L
\approx 0.1\mu$m may exhibit sliding columnar behavior.  However, in
the next round of experiments it will be important to prepare aligned
CL-DNA samples because powder-averaging complicates the functional
form of the scattering intensity and makes it difficult to identify
sliding columnar behavior.

We showed in Sec.~\ref{sec:sccorrel} that the density-density
correlation function $S_2({\bf r})$ [Eq.\ (\ref{S2r})]
displays different functional forms
depending on the magnitude of the in-plane separation ${\bf r}$.  The
crossover lengthscales for the correlation function are $l_x$, $l_z$,
$\xi_d$, $x^*$, and $z^*$.  The harmonic $2D$ smectic regime is defined
by $x,z < l_{x,z}$ and the nonlinear $2D$ smectic regime is defined by
$x,z > l_{x,z}$, where the nonlinear lengths $l_{x,z}$ are given in
Eq.\ (\ref{nonlinlengths}).  To estimate the nonlinear lengths, we must
determine $l_x$ and $l_z$ in terms the experimentally measured
quantities $d$ and the harmonic correlation length 
$\xi_z$ of Eq.\ (\ref{xi_z}). 
$l_x$ and $l_z$ can be written in terms of $d$ and $\xi_z$ by solving 
Eq.\ (\ref{xi_z}) for $B_2$ and using
\begin{equation}
K_2 = T\xi_p/2d, 
\end{equation}
where $\xi_p = 500$\AA~is the persistence length of DNA. 
Table~\ref{expnonlinlengths} evaluates the nonlinear lengths for
two DNA spacings of the experiments of references \cite{RadKol97,HTexp}.
It shows that $L < l_{x,z}$ and indicates that significant departure 
from harmonic 2D smectic behavior is not expected in agreement with these
experiments.

The finite length of the DNA molecules $l_{\rm DNA} \approx 16\mu$m
introduces another crossover lengthscale.  The density of DNA molecules
within a given layer is $\rho = 1/dl_{\rm DNA}$.
If we assume that each free end of a DNA strand corresponds to a
dislocation, then we can estimate the characteristic distance between
dislocations to be
\begin{equation}
\label{dislocationlength}
\xi_d = \sqrt{d l_{\rm DNA} } .
\end{equation}
This length is clearly smaller than the actual distance between
dislocations because not every end of a DNA strand has to produce a
dislocation: a series of DNA strands can align in a row to produce a layer
of a $2D$ smectic.  At lengths scales greater than $\xi_d$, the $2D$ smectic
behaves like a nematic\cite{Toner2d}.  Using the estimate of Eq.\
(\ref{dislocationlength}), we find $\xi_d
\approx 0.21\mu$m and $0.30\mu$m when $d=28$\AA~and $55$\AA,
respectively.  Note that $L < \xi_d$, and thus the subdomains are
small enough to possess 2D smectic ordering.  

\begin{table}[h]
\caption{The nonlinear lengths, $l_x$ and $l_z$, calculated 
as a function of the DNA spacing $d$ using the experimental values of the 
in-plane correlation length $\xi_z$.}
\begin{tabular}{ccc}
Nonlinear length & $d=28$\AA & $d=55$\AA\\
\tableline
$l_x$ & $0.35\mu$m & $0.24\mu$m\\
$l_z$ & $9\mu$m & $6\mu$m \\
\end{tabular}
\label{expnonlinlengths}
\end{table}

We can also estimate the
energy cost for creating hairpin edge dislocations within the 2D
smectic lattices.  Hairpins cause the DNA director to change by $\pi$
over a lattice spacing $d$.  The energy cost for a hairpin can be
estimated from the 2D bending energy.  If we ignore relaxation of the
$2D$ smectic layers, we estimate the bending energy to be 
\begin{equation}
\label{hairpin}
{E_{hp} \over 2T} = {\pi \over 4} {\xi_p \over d},
\end{equation}  
which implies that hairpins are favored on lengthscales greater than 
\begin{equation}
\label{hairpinlength}
\xi_{hp} = d \exp\left(E_{hp} \over 2T\right).
\end{equation} 
Since $\xi_{hp} \gg \xi_d > L$ throughout the experimental range in
$d$, hairpins are not important for the current set of 
experiments.  We do not yet have accurate estimates of
$x^*$ and $z^*$ since the value of the orientational rigidity $K_y$ is
unknown.  Scattering experiments will see sliding columnar 
behavior on lengthscales less than $\xi_d$ if $x^*,z^* < \xi_d$.

\section{Discussion and Conclusion}
\label{sec:conclusion}
\setcounter{equation}{0}

In this paper, we have introduced the new sliding columnar (SC) phase
of matter which may exist in layered systems composed of
weakly-coupled 2D smectic lattices.  The sliding columnar phase is
characterized by weak positional but strong orientational correlations
between neighboring 2D smectic lattices.  The SC harmonic free energy
contains an orientational rigidity that aligns neighboring 2D smectic
lattices in addition to in-plane compression and bending moduli.  
The SC phase is characterized by a vanishing shear modulus for relative
displacements of $2D$ smectic lattices. The
presence of the orientational rigidity fundamentally alters the energy
spectrum.  In light of this, we have calculated the structural
properties of the sliding columnar phase such as the SC
displacement correlation function, scattering intensity, and
dislocation energy.  

Experimental research on layered liquid crystalline systems
studied in this work is in progress\cite{RadKol97}-\cite{SalKol98}. 
Correlation peaks theoretically discussed here have 
been observed in a number of scattering experiments 
on DNA-cationic lipid complexes.  Reference \cite{LTexp} reports 
DNA scattering patterns that clearly
reflect the short-range centered-rectangular order
depicted in Fig.\ \ref{fig:phase-xray}.
Reference  \cite{HTexp},
on the other hand, reports peaks in a different system at the simple
rectangular lattice positions $(0,0, \pm q_0)$ [i.e., at $(0,1)$ peaks].
In spite of this difference in 
scattering data,
it is likely that {\it both} systems have
short-range centered rectangular order in {\it real space}. 
Indeed, in a strongly fluctuating 
disordered systems, positions of
correlation peaks in $q$-space may not always reflect
length scales and short-range order in real
space. For example, 
the correlation peak in $q$-space in random
microemulsions\cite{micro1,gompschick} may occur at $q=0$
even in situations in which real space data still
show a finite structural length scale.
As discussed at the end of Sec.\ \ref{sec:densitycorr}, for sufficiently
small coupling 
inter-layer positional coupling $V^u/T$,
the scattering form factor $I_{DNA}({\bf q})$
exhibits a maximum at the simple
rectangular-lattice points $(0,0,\pm q_0)$ even though
the system has short-range centered rectangular order in 
real space.  The differences between the positions of the 
scattering peaks in Refs.\ \cite{HTexp} and \cite{LTexp} may be explained
by the differences in thermal disorder and/or inter-layer couplings
strengths of the two systems.  The interlayer
correlation length reported in Ref.\ \cite{LTexp} is several times longer
than that reported in Ref.\ \cite{HTexp}.  The former system
is stiffer and more ordered than the latter. 
Consequently the $(1, 1)$  and $(-1,1)$ correlation peaks in Ref.\
\cite{LTexp} reflect the centered-rectangular order in real space depicted
in Fig.\ \ref{scfig}, whereas the the $(0,1)$ peak reported in Ref.\
\cite{HTexp} reflects the merging of $(\pm 1, 1)$ peaks brought about by
strong thermal fluctuations and/or sufficiently weak inter-layer couplings. 

After completing this work, we learned from Joachim R\"{a}dler
of experiments \cite{private} showing a continuous variation of the DNA form
factor with changing strength of inter-layer coupling $V^{u}$
consistent with the change we describe above and in Sec.\ \ref{x-raysc} 
[crossover from centered rectangular scattering pattern to one resembling
that of a simple rectangular lattice].   In these experiments, the variation of
inter-layer positional coupling is produced by varying the period of
DNA lattices in galleries. This coupling generally increases with increasing
period of DNA lattices, most likely because the Coulomb interaction
increases and the elastic membrane deformation
increases, as suggested by electron density
maps of the system\cite{private}.
\acknowledgments

LG acknowledges support by Mylar Pharmaceuticals, Inc., and CSO and TCL
acknowledge support from the National Science Foundation
under grant DMR97--30405.  We are grateful for helpful discussions with
Ilya Koltover, Joachim R\"{a}dler, Cyrus Safinya, Tim Salditt, and John
Toner.

\appendix

\section{Derivation of Effective Hamiltonians}
\label{app:Heff}

In this appendix, we will derive the effective Hamiltonians, ${\cal H}^{SC}$
[Eq.\ (\ref{HSC})], ${\cal H}_z^{SC}$ [Eq.\ (\ref{scenergy})], ${\cal H}_h$ [Eq.\
(\ref{Hh})] obtained, respectively, by integrating out $h^n ({\bf r})$, $h^n (
{\bf r} )$ and $u_y^n ( {\bf r} )$, and $u_z^n ( {\bf r} )$ and $u_y^n ( {\bf r} )$ from
${\tilde {\cal H}} = {\cal H}^{\rm bend} + {\cal H}^{\rm com} + {\cal H}^{\rm rot}$, which we
can express in linearized form in Fourier space as
\begin{eqnarray}
{\tilde {\cal H}} & = & {1 \over 2} \int {d^3 q \over (2 \pi )^3} \{ A_h ( {\bf q} )
|h ( {\bf q} )|^2 + \sum_{\sigma} A_{\sigma} ( {\bf q} ) |u_{\sigma} ( {\bf q} ) |^2
\nonumber \\
& & - \sum_{\sigma} [\lambda_{\sigma} ( {\bf q} ) u_{\sigma} ( {\bf q} ) h( - {\bf q} )
+ \lambda_{\sigma} ( - {\bf q} ) u_{\sigma} ( - {\bf q} ) h( {\bf q} ) ]\} ,
\end{eqnarray}
where $\sigma = y,z$, and
\begin{eqnarray}
A_h ( {\bf q} ) & = & K_{3d} q_{\perp}^4 + 4 ( B_{3d}/a^2 ) \nonumber \\
A_y ( {\bf q} ) & = & 4 (B_{3d}/a^2) + K_{2d} q_x^4 \nonumber \\
A_z ( {\bf q} ) & = & B_{2d} q_z^2 + K_{2d} q_x^4 + K_y q_x^2 q_y^2 p(q_y a)
\nonumber \\
\lambda_y ( {\bf q} ) & = & (2 B_{3d}/a^2) \left( 1 + e^{-i q_y a } \right)
\nonumber \\
\lambda_z ( {\bf q} ) & = & -(B_{uh}/a)i q_z \left(e^{-i q_y a} -1 \right) .
\end{eqnarray} 
Integration over $h({\bf q})$ yields 
\begin{eqnarray}
\label{tHSC}
{\tilde {\cal H}}^{SC} & = & {1 \over 2}\int{d^3 q\over (2 \pi)^3} [A_y' ( {\bf q} )
|u_y ({\bf q} )|^2 + A_z' |u_z ({\bf q} )|^2 \nonumber \\
& &- \mu( {\bf q} ) u_y ( {\bf q} ) u_z ( - {\bf q} ) - 
\mu( - {\bf q} ) u_y ( -{\bf q} ) u_z ( {\bf q} ) ] ,
\end{eqnarray}
where
\begin{eqnarray}
A_{\sigma}^{\prime} ( {\bf q} ) & = & A_{\sigma} ( {\bf q} ) - 
{|\lambda_{\sigma} ( {\bf q} )|^2 \over A_h ( {\bf q} )} \\
\mu ( {\bf q} ) & = & {\lambda_y ( {\bf q} ) \lambda_z ( - {\bf q} ) \over A_h ( {\bf q} )
} .
\end{eqnarray}
In the small ${\bf q}_{\perp}$ limit, these expressions reduce to
\begin{eqnarray}
A_y^{\prime}( {\bf q} ) & = & (B_{3d}/a^2)2( 1 - \cos q_y a) + K^{\alpha \beta}
q_{\alpha}^2 q_{\beta}^2 \\
A_z^{\prime} ( {\bf q} ) 
& = & B_{2d} q_z^2 - {B_{uh}^2 \over 4 B_{3d}}  q_z^2 
2( 1 - \cos q_y a) + K_{2d} q_x^4 \nonumber\\
& & + K_y q_x^2 q_y^2 p(q_y a) \\
\mu ( {\bf q} ) & = & -(B_{uh} /a) q_z \sin q_y a ,
\end{eqnarray}
where $K^{\alpha \beta}$ is the bending rigidity tensor of Eq.\
(\ref{Kab}).  Fourier transformation of Eq.\ (\ref{tHSC}) back to real
space using the low-${\bf q}_{\perp}$ form of $A_y ( {\bf q} )$, $A_z ( {\bf q} )$ and
$\mu ( {\bf q} )$ produces the sliding phase Hamiltonian ${\cal H}^{SC}$ of Eq.\
(\ref{HSC}). 

The effective Hamiltonian for $u_z$ is obtained by integrating ${\tilde
{\cal H}}^{SC}$ over $u_y$.  The result is
\begin{equation}
{\tilde {\cal H}}_z^{SC} = {1 \over 2} \int {d^3 q \over (2 \pi )^3 }
A_z^{\prime\prime} ( {\bf q} ) |u_z ( {\bf q} ) |^2 ,
\end{equation}
where
\begin{equation}
A_z^{\prime\prime} ( {\bf q} ) = A_z^{\prime} ( {\bf q} ) - { |\mu ( {\bf q} ) |^2
\over A_y^{\prime} ( {\bf q} ) } .
\end{equation}
In the ${\bf q}_{\perp} \rightarrow 0$ limit, $A_z^{\prime\prime} ( {\bf q} )$
becomes
\begin{eqnarray}
A_z^{\prime\prime} ( {\bf q} ) & \rightarrow & q_z^2 \Big[ B_{2d} - { B_{uh}^2
\over 4 B_{3d} } 2 ( 1 - \cos q_y a )  \nonumber \\
& & \qquad - {B_{uh}^2 \over B_{3d}} {\sin^2 q_y a \over 2 ( 1 - \cos qy a
)}\Big]  \\
& & + K_y q_x^2 q_y^2 p( q_y a ) + K_{2d} q_x^4\nonumber \\
& = & q_z^2 \left( B_{2d} - {B_{uh}^2 \over B_{3d}} \right) + K_{2d} q_x^4
, \label{A11} \nonumber \\
& & + K_y q_x^2 q_y^2 p( q_y a )  
\end{eqnarray}
where we used the identity 
\begin{equation}
1- \cos \phi + {\sin^2 \phi \over 1 - \cos \phi} = 2  
\end{equation}
to obtain Eq.\ (\ref{A11})
Transformation of ${\tilde {\cal H}}_z^{SC}$ to real space using the low
${\bf q}_{\perp}$ form of $A_z^{\prime\prime}$ produces ${\cal H}_z^{SC}$ of Eq.\
(\ref{scenergy}).  The absence of any $q_y$ dependence in the coefficient
of $q_z^2$ in $A_z^{\prime\prime} ( {\bf q} )$ can be traced the equality
$B_{uh}^2/B_{3d} = 4B_{zz}$ [Eq.\ (\ref{Bzz})]
that results from integrating out $h^n$ from
the original model. 
Effective Hamiltonians for $h^n ( {\bf r} )$ and $u_y^n ( {\bf r} )$ can be
obtained by respectively integrating out $u_y^n$ and $u_z^n$ from ${\cal H}^{\rm
tot}$ and $u_z^n$ from ${\cal H}^{SC}$.  The results are
\begin{eqnarray}
{\cal H}_h^{SC} & =  & {1 \over 2} \int {d^3 q \over (2 \pi )^3 } 
A_h^{\prime\prime} ( {\bf q} )|h( {\bf q} ) |^2 \\
{\cal H}_y^{SC} & = &  {1 \over 2} \int {d^3 q \over ( 2 \pi )^3 }  
A_y^{\prime\prime} ( {\bf q} ) |u_y ( {\bf q} ) |^2 .
\end{eqnarray}
where
\begin{eqnarray}
A_h^{\prime\prime} ( {\bf q} ) & = & 
A_h ( {\bf q} ) -
{|\lambda_z ( {\bf q} )|^2 \over A_z ( {\bf q} ) } - {|\lambda_y ( {\bf q} )|^2 \over
A_y ( {\bf q} ) }  \\
A_y^{\prime\prime} ( {\bf q} ) & = &
A_y ( {\bf q} ) - {|\lambda_y ( {\bf q} ) |^2 \over A_h ( {\bf q} ) - |\lambda_z ( {\bf q}
)|^2/A_z ( {\bf q} )} .
\end{eqnarray}
In the limit ${\bf q}_{\perp} \rightarrow 0$ with $q_x^2 \ll
(B_{2d}a^2/K_y)q_z^2$,
both of these Hamiltonian reduce to ${\cal H}_h$ in Eq.\ (\ref{Hh}).

\section{Calculation of $\lowercase{\langle (u_z^n)^2\rangle}$}
\label{app:usquared}

In this Appendix, the expression for the sliding columnar displacement
fluctuations given in (\ref{uflucts}) is derived.  To do this,
we evaluate the integral of the the Fourier transformed SC
correlator $G_{zz}({\bf q})$ over all $q$-space:
\begin{eqnarray}
\label{usquaredef}
\langle (u_z^n)^2 \rangle & = & \int {d^3q \over (2\pi)^3} {T
\over B q_z^2 + Kq_x^4 + K_y q_x^2 q_y^2 p(q_y a)},
\end{eqnarray}
where $p(u) = 2(1-\cos u)/u^2$.  The fluctuations diverge
at small wavenumbers ${\bf q} \sim 1/L$, where $L$ is the system size.
To calculate how the fluctuations scale with $L_x$, we set $L_z
\rightarrow \infty$ and $L_y \sim L_x$.  Note that the SC form for
$G_{zz}({\bf q})$ is valid only when $L_x \gg x^*$, where $x^* = a/\mu_y$
and $\mu_y = \sqrt{K_y/K}$. The first step in the calculation is to
perform the integration over $q_z$ with $\Lambda_z \rightarrow \infty$
and then use the fact that $G_{zz}({\bf q})$ is an even function of ${\bf
q}$ so that the remaining integrals run over only positive $q_x$ and
$q_y$.  Note that taking the $\Lambda_z \rightarrow \infty$ limit does
not alter $q_{x,y} \rightarrow 0$ divergences.  The resulting
expression
\begin{eqnarray} 
\label{stepone}
\langle (u_z^n)^2 \rangle & = &
{T \over 2 \pi^2 \sqrt{BK}} \int^{\Lambda_x}_
{L_x^{-1}} {dq_x \over q_x} \int^{\Lambda_y}_{L_y^{-1}} dq_y {1 \over
\sqrt{ q_x^2 + \mu_y^2 q_y^2 p(q_ya)} }, \nonumber
\end{eqnarray}  
where $\Lambda_y = \pi/a$, is made dimensionless by changing variables
to $v = q_x x^*$ and $w = q_y a$.  The $\ln^2L_x$ divergence of the
displacement fluctuations can be seen immediately by looking at the
$q_{x,y} \rightarrow 0$ limit of this expression.

The SC displacement fluctuations can be written as the sum of a continuum
term $I_c$ that does not depend on $p(w)$ and discrete term $I_d$
that does depend on $p(w)$.  
\begin{equation}
\label{contanddis}
\langle (u_z^n)^2 \rangle \equiv {T \over 2\pi^2 \sqrt{B K_y}} 
\left[I_c + I_d\right],
\end{equation}
where $I_c$ and $I_d$ are defined by 
\begin{eqnarray}
\label{gs}
I_c & = & \int^{\Lambda_x x^*}_{x^* L^{-1}_x} {dv \over v} 
\int^{a\Lambda_y}_{a L^{-1}_y} {dw \over \sqrt{v^2 + w^2}} \nonumber \\
I_d & = & \int^{\Lambda_x x^*}_{x^* L^{-1}_x} {dv \over v} 
\int^{a\Lambda_y}_{a L^{-1}_y} dw \left[ 
{1 \over \sqrt{v^2 + w^2 p(w)}} - {1 \over \sqrt{v^2 + w^2}} \right].
\nonumber 
\end{eqnarray}
We first focus on the continuum contribution $I_c$.  The 
integral over $w$ is straightforward;
\begin{eqnarray}
\label{steptwo}
I_c & = & \int^{\Lambda_x x^*}_{x^*L_x^{-1}} {dv \over v} \left[
f(\Lambda_y a, v) - f(L^{-1}_y a,v)\right] \\
& \equiv & I_c^{(1)} - I_c^{(2)}, \nonumber
\end{eqnarray}   
where 
\begin{equation}
\label{fdefine}
f(x,v) = \ln\left[ x + \sqrt{x^2 + v^2}\right].
\end{equation}
The integral over $v$ in $I_c^{(1)}$ can be evaluated by 
separating the function 
\begin{eqnarray}
f(\Lambda_y a, v) & = & \ln[2 \Lambda_y a] + 
\ln\left[{1\over 2} + {1 \over 2}\sqrt{1 + (v/\Lambda_y a)^2}\right], \nonumber
\end{eqnarray}
into a constant term and a term that is well-behaved at small $v$.  
We then insert this expression into (\ref{steptwo}) and find that
\begin{eqnarray}
\label{gc1}
I_c^{(1)} & = & \ln[2\Lambda_ya] \ln[\Lambda_x L_x] \\
& + & 
\int^{\Lambda_x x^*}_{x^* L_x^{-1}} {dv \over v} 
\ln\left[{1\over 2} + {1 \over 2}\sqrt{1 + (v/\Lambda_y a)^2}\right]. \nonumber
\end{eqnarray}
The first term diverges with system size $L_x$, and the second term is
nondivergent. $f(aL^{-1}_y,v)$ can also be separated into a constant
term and a term that depends on $v$.  We then integrate
$f(aL^{-1}_y,v)$ over $v$ to find
\begin{eqnarray}
\label{stepfour}
I_c^{(2)} & = & \ln[2a L_y^{-1}]\ln[\Lambda_x L_x] \\
& + & 
\int^{\Lambda_x x^*}_{x^*L_x^{-1}} {dv \over v} \ln\left[ {1 \over 2} +
{1 \over 2}\sqrt{1 + \left( {L_y v \over a} \right)^2 }\right]. \nonumber
\end{eqnarray}
The $L_y$ dependence in the integrand of the second term can be moved
to limits of the integral by changing variables to $s = L_y v/a$.  In
contrast to the previous expression for $I_c^{(1)}$ in (\ref{gc1}),
the large $s$ part of the integral in (\ref{stepfour}) diverges with
system size.  The divergence can be isolated by adding and subtracting
$\ln[s/2]/s$.  The resulting expression,
\begin{eqnarray}
\label{stepfive}
I_c^{(2)} & = & \ln[2a L_y^{-1}]\ln[\Lambda_x L_x] + {1 \over 2} \ln^2\left[
{\Lambda_x L_y \over 2 \mu_y}\right] -{1\over 2}\ln^2\left[ {L_y \over 
2 \mu_y L_x}\right] \nonumber \\
& + & \int_{L_y/L_x \mu_y}^{\Lambda_x L_y/\mu_y}
{ds \over s} \left( \ln\left[ {1 \over 2} + {1\over 2}\sqrt{1+s^2}\right]
- \ln[s/2] \right), \nonumber
\end{eqnarray}
has two terms that diverge and two terms that do not diverge with
system size.  Note that $L_y/L_x \mu_y$ is ${\cal O}(1)$ since $L_x \sim
L_y$.  We then subtract $I_c^{(2)}$ from $I_c^{(1)}$, drop the
nondivergent terms, and set $L_y = L_x$ to obtain
\begin{eqnarray}
\label{gccollect}
I_c = \ln[\Lambda_x L_x] \ln[\Lambda_y L_y] - {1 \over 2} \ln^2\left[
{\Lambda_x L_y \over 2 \mu_y}\right] = {1 \over 2} 
\ln^2[2 \mu_y \Lambda_y L_x] \nonumber
\end{eqnarray}  
for the continuum contribution to the displacement fluctuations.

The discrete contribution is obtained by evaluating
\begin{equation}
\label{discretedef}
I_d = \int^{\Lambda_x x^*}_{x^* L_x^{-1}} {dv \over v} F(v),
\end{equation}
where 
\begin{equation}
\label{Fdef}
F(v) = \int_0^{\pi} dw \left[ {1 \over \sqrt{v^2 + w^2 p(w)} }
- {1 \over \sqrt{v^2 + w^2} } \right].
\end{equation}
In the definition of $F(v)$, we can take the lower limit to zero 
since the small $w$ part of integral is well-behaved.  In
contrast to the continuum term, the discrete term diverges
logarithmically with system size.  To see this, we expand $F(v)$ around
$v=0$ and find $F(v) \approx F(0) + a v^2 + {\cal O}(v^4)$, where
$F(0) = \ln[4/\pi]$ and $a$ is a constant.  Thus, $I_d =
\ln[4/\pi]\ln[\Lambda_x L_x]$ plus terms that do not diverge with
system size.  To make the argument of the $\ln$ term in $I_d$ match the
$\ln^2$ term in $I_c$, we add and subtract the constant
$\ln[4/\pi]\ln[2\mu_y \Lambda_y/\Lambda_x]$ to find
\begin{equation}
\label{discrete}
I_d = \ln[4/\pi]\ln[2 \mu_y \Lambda_y L_x].
\end{equation}  
We then add $I_c$ and $I_d$ and find the following
expression for the displacement fluctuations in the limit $L_z \rightarrow
\infty$ and $L_x \sim L_y$:
\begin{equation}
\langle (u_z^n)^2 \rangle = l_u^2 \ln^2 \left[8 L_x/x^* \right].
\end{equation}

\section{Calculation of $\lowercase{g^{(1)} (na)}$} 
\label{app:g1}

In this appendix, we calculate the divergent part of the position
correlation function $g^{(1)}(na)$ in (\ref{g1def}).  We 
perform the $q_z$ integration, switch to dimensionless variables, and  
find
\begin{eqnarray}
\label{diverge}
g^{(1)}(na) = 2 l_u^2 \left[ \int^{\Lambda_x x^*}_{x^* L_x^{-1}}
{dt \over t} S_n(t) + {\overline A}_n(\Lambda_x,\Lambda_z) \right],
\end{eqnarray}
where $S_n(t)$ was defined previously in (\ref{sndef}) and  
\begin{eqnarray}
\label{adef}
{\overline A}_n(\Lambda_x,\Lambda_z) & = & 
- \int_0^{\Lambda_x x^*} {dt \over t}
\int_0^{\pi} du {1-\cos nu \over \sqrt{t^2 + u^2 p(u)}} \\
& \times & \left(
1 - {2 \over \pi} \arctan \left[ {\Lambda_z z^* \over t\sqrt{t^2 + u^2 p(u)}}
\right] \right). \nonumber
\end{eqnarray}
Note that the integral defining ${\overline A}_n(\Lambda_x,\Lambda_z)$
does not have an infrared divergence as $t \rightarrow 0$ ($L_x
\rightarrow
\infty$), and thus it is a well-defined number.  In the final step,
we isolate the $\ln L_x$ divergence in the first term of
(\ref{diverge}) to find
\begin{equation}
g^{(1)}(na) = 2 l_u^2 S_n(0) \ln \left[
A_n(\Lambda_x,\Lambda_z) L_x/x^* \right],
\end{equation}
where
\begin{equation}
A_n(\Lambda_x,\Lambda_z) = \exp[c^{(1)}_n + c^{(2)}_n(\Lambda_x) 
+ c^{(3)}_n(\Lambda_x,\Lambda_z)]
\end{equation}
depends on the layer index $n$ and the ultraviolet cutoffs with  
\begin{eqnarray}
c^{(1)}_n & = & \int_0^1 {dt \over t} \left[ {S_n(t) \over S_n(0)} 
- 1 \right],\label{C5} \\
c^{(2)}_n(\Lambda_x) & = & \int_1^{\Lambda_x x^*} {dt \over t} 
{S_n(t) \over S_n(0)} \label{C6}\\
c^{(3)}_n(\Lambda_x,\Lambda_z) & = & { {\overline A}_n(\Lambda_x,\Lambda_z) 
\over S_n(0)}.
\end{eqnarray}
Note that $A_n(\Lambda_x,\Lambda_z)$ is well-defined in the limit 
$\Lambda_{x,z} \rightarrow \infty$.

\section{Calculation $\lowercase{g^{(2)} ( {\bf r} )}$}
\label{app:corrfunction}

In this Appendix, we evaluate the SC position correlation function
\begin{equation}
g^{(2)}({\bf r}) = {1 \over 2} \langle [u_z^0({\bf r})-
u_z^0(0)]^2\rangle
\end{equation} 
between two DNA strands located in layer $n=0$ and
separated by ${\bf r}$ in the $xz$ plane.  For general separations,
$g^{(2)}({\bf r})$ cannot be expressed in closed form.  The aim of this
Appendix is to calculate $g^{(2)}({\bf r})$ along the special directions
$z=0$, $x \gg x^*$ and $x=0$, $z \gg z^*$.

\subsection{Large $x$, Small $z$ Limit}
\label{sec:xlimit}

The following expression for $g^{(2)}(x,0)$ is obtained by setting $z$ 
to zero in (\ref{g2r}):
\begin{equation}
\label{corrstepone}
g^{(2)}(x,0) = T \int {d^3 q \over (2\pi)^3} {1-\cos(q_x x) 
\over B q_z^2 + Kq_x^4 + K_y q_x^2 q_y^2 p(q_y a)}.
\end{equation}
The first step in the derivation of $g^{(2)}(x,0)$ is to perform the
integration over $q_z$ with $\Lambda_z \rightarrow \infty$.  The $q_z$
integration yields
\begin{equation}
\label{gxstepone}
g^{(2)}(x,0) = {T \over 2\pi^2 \sqrt{B K_y} } I(x,\Lambda_x),
\end{equation}
where
\begin{eqnarray}
\label{gxsteptwo}
I(x,\Lambda_x) & = & \int_0^{\Lambda_x} dq_x {1 - \cos(q_x x) \over q_x} 
\int_0^{\pi/x^*} {dq_y \over \sqrt{q_x^2 + q_y^2 p(q_yx^*)} }. \nonumber
\end{eqnarray}
We then decompose $I(x,\Lambda_x) \equiv I_c(x,\Lambda_x) 
+ I_d(x,\Lambda_x)$ into continuum and discrete contributions 
as we did previously in Appendix~\ref{app:usquared}, where  
\begin{eqnarray}
\label{Icontinuum}
I_c(x,\Lambda_x) & = &\int_0^{\Lambda_x} dq_x {1- \cos(q_x x) \over q_x} 
\int_0^{\pi/x^*} {dq_y \over \sqrt{q_x^2 + q_y^2}} \nonumber \\
\label{Idiscrete}
I_d(x,\Lambda_x) & = & \int_0^{\Lambda_x} dq_x {1- \cos(q_x x) \over q_x} 
\int_0^{\pi/x^*} dq_y \nonumber \\
& & \left[ {1 \over \sqrt{q_x^2 + q_y^2 p(q_yx^*)} }
- {1 \over \sqrt{q_x^2 + q_y^2} } \right]. \nonumber
\end{eqnarray}
Since the $\Lambda_x \rightarrow \infty$ limit is well-defined, we
calculate $I_c(x) \equiv I_c(x,\infty)$ and $I_d(x) \equiv I_d(x,\infty)$
and drop terms that depend on the finite ultraviolet cutoff.

To calculate the continuum contribution, we first set $q_x = u q_y$ and 
then $v = q_y x$.  These changes of variables yield
\begin{equation}
\label{continuumcont}
I_c(x) = \int_0^{\pi x/x^*} {dv \over v} K(v),
\end{equation}
where
\begin{equation}
K(v) = \int_0^{\infty} du {1-\cos(uv) \over u\sqrt{1 + u^2}}.
\end{equation}
The strategy for calculating the $\ln^2x$ term in $I_c(x)$ is to isolate the
part of $K(v)$ that scales as $\ln v$ for large $v$.  To this end, we
write
\begin{eqnarray}
\label{K(v)}
K(v) & = & \int_0^1 {du \over u} [1-\cos(uv)] 
+ \int_1^{\infty} {du \over u} {1-\cos(uv) \over \sqrt{1+u^2} } \nonumber \\
& + & \int_0^1 {du \over u} \left[1- \cos(uv)\right]
\left[ {1 \over \sqrt{1+u^2}} - 1\right]. 
\end{eqnarray}
It is obvious that only the first term has the correct scaling; 
the remaining terms in $K(v)$ are then separated into constants
and functions of $v$ that are well-behaved either as $v \rightarrow 0$ 
or $v \rightarrow \infty$.  This partitioning leads to
\begin{equation}
K(v) = \ln(B v) + {\widetilde K}(v),
\end{equation}
where $B = 2e^{\gamma}$, $\gamma$ is Euler's constant, and 
\begin{eqnarray}
\label{tK(v)}
{\widetilde K}(v) & = & \int_v^{\infty} du {\cos u \over u} - \int_1^{\infty} 
{du \over u} {\cos(uv) \over \sqrt{1+u^2} } \nonumber \\
& - & \int_0^1 du {\cos(uv) \over u} \left[{1 \over \sqrt{1+u^2}} -1\right]  
\end{eqnarray}
scales as $1/v^2$ for large $v$.

We then insert $K(v)$ into (\ref{continuumcont}) and break the 
integral over $v$ into small- and large-$v$ parts to obtain
\begin{eqnarray}
\label{finalcontstep}
I_c(x) & = & \int_0^1 {dv \over v} K(v) + \int^{\pi x/x^*}_1 {dv \over v}
\ln[B v]  \\
& + & \int^{\pi x/x^*}_1 {dv \over v} {\widetilde K}(v). \nonumber
\end{eqnarray}
Next, we evaluate the integral over $v$ in the second term, collect constants,
and find
\begin{equation}
I_c(x) = {1\over 2} \ln^2[2 e^{\gamma} \pi x/x^*] + A_x, 
\end{equation}
where 
\begin{equation}
\label{A_x}
A_x = -{1\over 2}\ln^2[2e^{\gamma}] + \int_0^1 {dv \over v} K(v)
+ \int_1^{\infty} {dv \over v} {\widetilde K}(v). 
\end{equation}
The second and third terms in $A_x$ are finite since $K(v)$ scales as
$v^2$ for small $v$ in the former and there is phase cancellation from the
$\cos(uv)$ factor at large $v$ in the later.

We will now calculate the discrete contribution to $g^{(2)}(x,0)$.  The 
first step is to rewrite $I_d(x)$ in dimensionless form:
\begin{equation}
\label{gddimensionless}
I_d(x) = \int_0^{\Lambda_x x^*} {dv \over v} [1-\cos(v x/x^*)]
F(v),
\end{equation} 
where $F(v)$ was defined previously in (\ref{Fdef}).  We next 
break the integral over $v$ into small- and large-$v$ parts and 
take the $x\gg x^*$ and $\Lambda_x \rightarrow \infty$ limits
to obtain
\begin{eqnarray}
\label{runningout}
I_d(x) & = & F(0) \int_0^1 {dv \over v} [1-\cos(v x/x^*)] + 
\int_1^{\infty} {dv \over v}F(v) \nonumber \\
& + & \int_0^1 {dv \over v} [F(v) - F(0)].
\end{eqnarray}  
Note that taking the $x\gg x^*$ limit removed the $\cos(v x/x^*)$ 
terms from the last two terms in (\ref{runningout}) due to phase 
cancellations.  It is again obvious that the first term 
scales logarithmically with $x/x^*$, and thus
\begin{equation}
I_d(x) = \ln[4/\pi]\ln\left[e^{\gamma} {x \over x^*}\right] + B_x,
\end{equation}
where
\begin{equation}
\label{B_x}
B_x = \int_0^1 {dv \over v} [F(v) - F(0)] + \int_1^{\infty} {dv \over v}F(v)
\end{equation}
is a constant.  The last step in the calculation of $g(x,0)$ is
to add the continuum and discrete terms, $I_c(x)$ and $I_d(x)$.
The final result is 
\begin{equation}
\label{g2x0}
g^{(2)}(x,0) = l_u^2 \left(
\ln^2\left[ 8 e^{\gamma}{x \over x^*}\right] + C_x\right),
\end{equation}
where $C_x = 2(A_x + B_x) - \ln^2[4/\pi] - 2\ln[4/\pi]\ln[2\pi]$.  
$A_x$ and $B_x$ are computed later in this appendix, see Eqs.\ (\ref{A_x1})
and (\ref{B_x1}).  By using these equations, we eventually find that $C_x$
in Eq.\ (\ref{g2x0}) vanishes; $C_x = 0$. 
\subsection{Large $z$, Small $x$ Limit}
\label{sec:zlimit}

The calculation of $g^{(2)}(0,z)$ is similar to the calculation of
$g^{(2)}(x,0)$.  The expression for $g^{(2)}(0,z)$ is obtained by setting $x$
to zero in (\ref{g2r}).  The first step in the
calculation is to perform the integration over $q_z$ with $\Lambda_z
\rightarrow
\infty$ which yields 
\begin{equation}
g^{(2)}(0,z) = {T \over 2\pi^2 \sqrt{B K_y}}I(z,\Lambda_x),
\end{equation}
where 
\begin{eqnarray}
I(z,\Lambda_x) & = & \int_0^{\Lambda_x} {dq_x \over q_x} \int_0^{\pi/x^*} dq_y
{1 - e^{-z\lambda q_x \sqrt{q_x^2 + q_y^2p(q_y x^*)} } \over 
\sqrt{q_x^2 + q_y^2p(q_y x^*)} }. \nonumber
\end{eqnarray}
In what follows, we set $\Lambda_x \rightarrow
\infty$, drop terms that depend on the finite ultraviolet cutoff, and define
$I(z) \equiv I(z,\infty)$.  The second step is to change variables to
$u=q_x/q_y$ and $v = \lambda z q_y^2$ and decompose $I(z) \equiv
I_c(z) + I_d(z)$ into continuum and discrete terms, where
\begin{eqnarray}
\label{zcontinuum}
I_c(z) & = & {1\over 2} \int_0^{\tau} {dv \over v} \int^{\infty}_0 du
{1-e^{-vu\sqrt{1+u^2}} \over u\sqrt{1+u^2} }\\
I_d(z) & = & \int_0^{\infty} {dv \over v} \left[F(v,0) - F(v, vz/z^*)
\right],  
\end{eqnarray} 
with $\tau = \pi^2 z/z^*$, $z^* = a^2/\mu_y^2 \lambda$, 
\begin{equation}
F(v,\tau) = \int_0^{\pi} du \left[ { e^{-\tau\sqrt{v^2 + u^2p(u)}} \over 
\sqrt{v^2 + u^2p(u)} } - { e^{-\tau\sqrt{v^2 + u^2} } \over \sqrt{v^2 + 
u^2} } \right], 
\label{F(v,t)}
\end{equation}
and $F(v,0)$ is equivalent to $F(v)$ defined in Eq.\ (\ref{Fdef}).

We first focus on the continuum contribution to $g^{(2)}(z,0)$.
The integral over $v$ in $I_c(z)$ can be broken into small- and 
large-$v$ parts,
\begin{equation}
\label{Izfirst}
I_c(z) = {1 \over 2} \int_0^1 {dv \over v} J(v) + {1 \over 2} \int_1^{\pi^2
z/z^*} {dv \over v} J(v),
\end{equation}
where 
\begin{equation}
\label{J(v)}
J(v) = \int_0^{\infty} du {1 - e^{-vu\sqrt{1+u^2}} \over u\sqrt{1+u^2} }.
\end{equation}
The strategy is to extract the part of $J(v)$ that scales as 
$\ln v$ for large $v$.  If $J(v) \sim \ln v$ for large $v$, 
$I_c(z)$ will scale as $\ln^2[z/z^*]$ as expected.  Note that $J(v)$ scales 
as $v^2$ for small $v$, and thus the first term in (\ref{Izfirst}) is 
a finite constant.   After some algebra, we find  
\begin{equation}
\label{jvresult}
J(v) = \ln[D v] + {\widetilde J}(v),
\end{equation}
where $D = 2e^{\gamma}$,
\begin{eqnarray}
\label{tJ(v)}
{\widetilde J}(v) & = & \int^{\infty}_{\sqrt{2}v} du {e^{-u} \over u}
-\int^{\infty}_1 du {e^{-vu\sqrt{1+u^2}} \over u\sqrt{1+u^2} } \nonumber \\
& - & \int_0^1 du {1-\Phi(u) \over u\sqrt{1+u^2}} e^{-vu\sqrt{1+u^2} },
\end{eqnarray}
and
\begin{equation}
\label{Phi(u)}
\Phi(u) = { 1+2u^2 \over \sqrt{1+u^2} }.
\end{equation} 
We can now insert the expression for $J(v)$ in (\ref{jvresult}) into
(\ref{Izfirst}) and obtain the continuum contribution 
\begin{equation}
I_c(z) = {1\over 4}\ln^2\left[ 2e^{\gamma} \pi^2 {z \over z^*}\right]
+ A_z,
\end{equation}
where
\begin{eqnarray}
\label{A_z}
A_z & = & -{1 \over 4} \ln^2\left[2e^{\gamma}\right] + {1\over 2} \int_0^1 {dv 
\over v} J(v) \nonumber \\
& + & {1\over 2}\int_1^{\infty} {dv \over v} {\widetilde J}(v)
\end{eqnarray}
is a constant.  Note that ${\widetilde J}(v)$ decays exponentially 
for large $v$, and thus the third term in $A_z$ is finite.

We now concentrate on the discrete contribution to $g^{(2)}(z,0)$.
The integral over $v$ in $I_d(z) \equiv I^{(1)}_d + I^{(2)}_d$ can also 
be broken into small- and large-$v$ parts, where 
\begin{equation}
\label{I1dz}
I^{(1)}_d  =  \int_0^1 {dv \over v} \left[F(v,0) - F(v,vz/z^*)\right]
\end{equation}
and $I^{(2)}_d$ is an identical expression except the limits on the 
integral over $v$ run from one to infinity.  To isolate the $\ln z$ 
term in $I^{(1)}_d$, we change variables to $t = vz/z^*$ and take 
the $z \gg z^*$ limit.  In the large $z$ limit, (\ref{I1dz}) becomes
\begin{equation}
I_d^{(1)} = F(0,0) \ln\left[z \over z^*\right] + {\overline B}_z,
+ \int_0^1 {d v \over v} [F(v,0) - F(0,0)]
\end{equation}
where $F(0,0) = \ln[4/\pi]$ and  
\begin{equation}
{\overline B}_z = \int_0^1 {dt \over t} \left[F(0,0) - F(0,t)\right]
- \int_1^{\infty} {dt \over t} F(0,t)
\label{D30}
\end{equation}
is a constant.  The large-$v$ contribution to $I_d$, 
\begin{equation}
I_d^{(2)} = \int_1^{\infty} {dv \over v}F(v,0),
\end{equation}
is simply a constant when $z \gg z^*$.  We then collect the discrete
contributions and find
\begin{eqnarray}
I_d(z) & = & I_d^{(1)} + I_d^{(2)} , \nonumber \\
&= & \ln\left[4 \over \pi\right]\ln\left[z \over z^*\right] + B_z ,
\end{eqnarray}
where $B_z = {\overline B}_z + B_x$ with $B_x$ given by Eq.\ (\ref{B_x}).

The last step in the calculation of $g^{(2)}(0,z)$ is to add $I_c(z)$ 
and $I_d(z)$.  The final result is 
\begin{equation}
g^{(2)}(0,z) = l_u^2
\left({1 \over 2} \ln^2\left[32 e^{\gamma} {z \over z^*}\right] + C_z \right),
\end{equation}
where $C_z = 2(A_z + B_z) -2 \ln^2[4/\pi] -2
\ln[4/\pi]\ln[2e^{\gamma}\pi^2]$. $A_z$ and $B_z$ are calculated later in
this appendix [See Eqs.\ (\ref{A_z1}) and (\ref{B_z1})]. By using these
equations, we find $C_z= \pi^2/8$.

\subsection{Calculation of $A_{x}$, $B_{x}$, $A_{z}$, and $B_{z}$}
In Secs. 1 and 2 of this Appendix, we anticipated that the numerical constants
$A_{x}$,$B_{x}$,$A_{z}$, and $B_{z}$ have the values:
\begin{eqnarray}
\label{A_x1}
A_{x}&=&{ \pi^{2} \over 24} , \\
B_{x}& = & -{ \pi^{2} \over 24}+4 \ln ^{2}(2) \nonumber \\
& & - [\ln(2)][\ln(\pi)]-{\ln^{2}(\pi) \over 2} ,
\label{B_x1} \\
\label{A_z1}
A_{z}&=&{ 7 \pi^{2} \over 48} , \\
B_{z}& = & -{ \pi^{2} \over 12}+6 \ln^{2} (2)+ 2 \gamma \ln(2) \nonumber \\
& & -[\ln(2)][\ln(\pi)]-\gamma \ln(\pi)-\ln^{2}(\pi) .
\label{B_z1}
\end{eqnarray}
In this section we outline calculations yielding Eqs.\ (\ref{A_x1}) 
to (\ref{B_z1}).
We begin by deriving more explicit formulas for these constants.
Thus, for $A_{x}$, we obtain, by Eqs.\ (\ref{K(v)}), (\ref{tK(v)}), 
and (\ref{A_x}),
\begin{equation}
A_{x}=-{1\over 2}\ln^{2}(2e^{\gamma})+A_{x}^{(1)}+A_{x}^{(2)}
+\gamma A_{x}^{(3)} ,
\label{A_xn}
\end{equation}
with
\begin{eqnarray}
A_{x}^{(1)} & = & -\int_{0}^{1}dv \ln(v) {1-\cos(v) \over v} \nonumber \\
& & +\int_{1}^{\infty}dv \ln(v) {\cos(v) \over v} ,
\label{Ax12}
\end{eqnarray}
\begin{eqnarray}
A_{x}^{(2)}& = &\int_{0}^{1}{du \over u}
\bigg[{1\over \sqrt{ 1+u^{2}}}-1 \bigg] \ln(u) \nonumber \\
& & +\int_{1}^{\infty}{du \over u} {1\over \sqrt {1+u^{2}}} \ln(u) ,
\label{Ax22}
\end{eqnarray}
and
\begin{eqnarray}
A_{x}^{(3)}& = & \int_{0}^{1}{du \over u}\bigg[{1\over \sqrt{1+u^{2}}}-1 \bigg]
\nonumber \\
& & +\int_{1}^{\infty}{du \over u} {1\over \sqrt{ 1+u^{2}}} .
\label{A_x3}
\end{eqnarray}
Likewise, for the constant $A_{z}$, we obtain, by Eqs.\ (\ref{J(v)}), 
(\ref{tJ(v)}), (\ref{Phi(u)}), and (\ref{A_z}),
\begin{eqnarray}
A_{z} & = & -{1\over 4}\ln^{2}(2e^{\gamma}) \nonumber \\
& & +{1 \over 2}\big[A_{z}^{(1)}+{1\over 8}\ln^{2}(2)+
{\gamma \over 2}\ln(2)\big] \nonumber \\
& & +{1 \over 2} \big[A_{z}^{(2)}+\gamma A_{z}^{(3)}\big] , 
\label{D42}
\end{eqnarray}
where
\begin{eqnarray}
A_{z}^{(1)} & = & -\int_{0}^{1}dx \ln(x) {1-e^{-x} \over x} \nonumber \\
& & +\int_{1}^{\infty}dx \ln(x) {e^{-x} \over x} ,
\label{D43}\\
A_{z}^{(2)} & = &
\int_{0}^{1}du  {1-\Phi(u) \over u \sqrt {1+u^{2}}} \ln(u \sqrt {1+u^{2}})
\nonumber \\
& & +\int_{1}^{\infty}du {1 \over u \sqrt {1+u^{2}}}\ln(u \sqrt{ 1+u^{2}}),
\label{D44}
\\
A_{z}^{(3)} & = & \int_{0}^{1}du {1-\Phi(u) \over u \sqrt{ 1+u^{2}}}
\nonumber \\
& & +\int_{1}^{\infty}du {1 \over u \sqrt{ 1+u^{2}}} 
\label{A_z3}
\end{eqnarray}
with $\Phi(u)$ defined in Eq.\ (\ref{Phi(u)}). 
We proceed to compute the integrals in Eqs.\ (\ref{A_xn}) to (\ref{A_z3}). 
Integrals (\ref{A_x3}) and (\ref{A_z3}) can be done by elementary 
integration methods yielding
\begin{equation}
A_{x}^{(3)}=\ln(2), ~~~A_{z}^{(3)}={\ln(2) \over 2}. 
\label{D46}
\end{equation}
Other integrals are not elementary, and we outline their 
calculation in the following.
Thus, we compute the integrals $A_{x}^{(1)}$ and $A_{z}^{(1)}$ first by
showing that
\begin{equation}
A_{x}^{(1)}=A_{z}^{(1)}-{\pi^{2} \over 8} . 
\label{A_x11}
\end{equation}
and, next, by showing that
\begin{equation}
A_{z}^{(1)}={1\over 2}\left({d^{2}\Gamma \over dz^{2}}\right)_{z=1} , 
\label{A_zgamma}
\end{equation}
where $\Gamma(z)$ signifies the gamma function.

To obtain Eq.\ (\ref{A_x11}), consider the complex 
function $f(z)=e^{-z} \ln(z)/z$.
In the complex $z=x+iy$ plane, $f(z)$ is analytic inside 
the contour $C$ made of the
following four segments: $C_{1}:[z=x; \epsilon <x <R]$, 
$C_{2}:[z=R e^{i\theta}; 0<\theta<\pi/2]$,
$C_{3}:[z=iy; R>y>0]$, and $C_{4}:[z=\epsilon e^{i\theta}; \pi/2>\theta>0]$.
By applying the Cauchy residue theorem to the above $f(z)$ along the contour 
$C=C_{1}+C_{2}+C_{3}+C_{4}$ in the limit 
$\epsilon \rightarrow 0$ and $R \rightarrow \infty$,
we directly obtain the  relation
(\ref{A_x11}) between $A_{x}^{(1)}$ and $A_{z}^{(1)}$. 
Next, we demonstrate Eq.\ (\ref{A_zgamma}) by differentiating the 
standard integral representation of the gamma function. This yields
\begin{equation}
\left({d\Gamma \over dz}\right)_{z=1}=\int_{0}^{\infty}dx \ln(x) e^{-x}=-\gamma,
\end{equation}
\begin{equation}
\left({d^{2}\Gamma \over dz^{2}}\right)_{z=1}=
\int_{0}^{\infty}dx \ln^{2}(x) e^{-x} . 
\label{d2gamma}
\end{equation}
We then rewrite Eq.\ (\ref{d2gamma}) as,
\begin{eqnarray}
\left({d^{2}\Gamma \over dz^{2}}\right)_{z=1} & = &
\int_{0}^{1}dx (e^{-x}-1) \ln^{2}(x) \nonumber \\
& & +\int_{0}^{1}dx \ln^{2}(x)
+\int_{1}^{\infty}dx e^{-x}  \ln^{2}(x)
\end{eqnarray}
and integrate by parts the first integral 
[by writing $(e^{-x}-1) dx=d(1-e^{-x}-x)$,etc.]
as well as the last integral [by writing $e^{-x}dx=d(-e^{-x})$, etc.]. 
After few elementary integrations, this  yields our Eq.\ (\ref{A_zgamma})
which, combined with Eq.\ (\ref{A_x11}), yields values 
of the integrals $A_{x}^{(1)}$ and
$A_{z}^{(1)}$. To complete this calculation we need also the value of the
second derivative of $\Gamma(z)$ at $z=1$ that enters Eq.\ (\ref{A_zgamma}). 
To compute it, we use a relation from the theory of the gamma function:
\begin{equation}
{d\Gamma (z) \over dz}=\Gamma (z) \Psi(z) , 
\label{Dgamma}
\end{equation}
with
\begin{eqnarray}
\Psi(z) & = & -\gamma+\big(1-{1\over z}\big)+
\big({1\over 2}-{1\over z+1}\big) \nonumber \\
& & +\big({1\over 3}-{1\over z+2}\big)+... ;
\label{Psi}
\end{eqnarray}
$\Gamma(1)=1$, $(d \Gamma/dz)_{z=1}=\Psi (1)=-\gamma$.
By differentiating Eq.\ (\ref{Dgamma}),
\begin{eqnarray}
\left({d^{2}\Gamma \over dz^{2}}\right)_{z=1}& = &
\left({d\Gamma \over dz}\right)_{z=1}\Psi(1)
 +\Gamma (1)\left({d\Psi \over dz}\right)_{z=1}
\nonumber \\
& = & \gamma^{2}+\left({d\Psi \over dz}\right)_{z=1}.
\end{eqnarray}
The use of Eq.\ (\ref{Psi}) to compute $d\Psi(z)/dz$ at $z=1$  yields
\begin{eqnarray}
\left({d^{2}\Gamma \over dz^{2}}\right)_{z=1} & = &
\gamma^{2}+1+{1 \over 2^{2}}+{1\over 3^{2}}+{1\over 4^{2}}+...
\nonumber \\
&= & \gamma^{2}+{\pi^{2} \over 6} .
\label{D2gamma}
\end{eqnarray}
By Eqs.\ (\ref{D2gamma}), (\ref{A_x11}), and (\ref{A_zgamma}), 
we finally obtain
\begin{equation}
A_{x}^{(1)}={\gamma^{2} \over 2}-{\pi^{2} \over 24}, 
~~A_{z}^{(1)}={\gamma^{2} \over 2}+{\pi^{2} \over 12} .
\label{D56}
\end{equation}

Next, we proceed to compute the integral $A_{x}^{(2)}$ in Eq.\ (\ref{Ax22}). 
For this purpose, consider the integrals
\begin{eqnarray}
A_{x}^{(2)}(t)& = &\int_{0}^{t}{du \over u}
\bigg[{1\over \sqrt{ 1+u^{2}}}-1 \bigg] \ln(u) \nonumber \\
& & +\int_{t}^{\infty}{du \over u} {1\over \sqrt{ 1+u^{2}}} \ln(u) 
\label{D57}\\
A_{x}^{(3)}(t)& = &\int_{0}^{t}{du \over u}
\bigg[{1\over \sqrt {1+u^{2}}}-1 \bigg] \nonumber \\
& & +\int_{t}^{\infty}{du \over u} {1\over \sqrt{ 1+u^{2}}}
\label{D58}
\end{eqnarray}
For $t=1$, $A_{x}^{(2)}(t=1)=A_{x}^{(2)}$ is the desired integral,
whereas $A_{x}^{(3)}(t=1)=A_{x}^{(3)}=\ln(2)$, by (\ref{D46}). Further,
by Eqs.\ (\ref{D57}) and (\ref{D58}), $d A_{x}^{(2)}(t)/dt=-\ln(t)/t$, and 
$dA_{x}^{(3)}(t)/dt=-1/t$. By integrating these relations over $t$,
\begin{eqnarray}
A_{x}^{(2)}&=&A_{x}^{(2)}(t)+{\ln^{2}(t) \over 2}, 
\label{D59}\\
A_{x}^{(3)}& = &A_{x}^{(3)}(t)+\ln(t), 
\label{D60}
\end{eqnarray}
for {\it any} $t>0$. By Eqs.\ (\ref{D57}) to (\ref{D60}), 
with $t=\epsilon \rightarrow 0$,
\begin{eqnarray}
A_{x}^{(2)}& = &
\lim_{\epsilon \rightarrow 0}
\left[\int_{\epsilon}^{\infty}{du \over u} {1\over \sqrt{ 1+u^{2}}}\ln(u)
+{\ln^{2}(\epsilon) \over 2} \right] \label{D61} \\
A_{x}^{(3)} & = &
\lim_{\epsilon \rightarrow 0}
\left[
\int_{\epsilon}^{\infty}{du \over u} {1\over \sqrt{ 1+u^{2}}}
+\ln(\epsilon) \right]
\label{D62}
\end{eqnarray}
To proceed, it is useful to form the difference
\begin{equation}
\Delta=A_{x}^{(2)}-{1\over 2}[A_{x}^{(3)}]^{2}. 
\label{D63}
\end{equation}
Using here Eqs.\ (\ref{D61}) and (\ref{D62}), we find
\begin{equation}
\Delta=\int_{0}^{\infty}du {1 \over u \sqrt {1+u^{2}}}
\int_{0}^{u}{dv \over v}\left[ 1-{1 \over \sqrt{ 1+v^{2}}}\right] . 
\label{D64}
\end{equation} 
The integral over $v$ here can be done by  changing
$v \rightarrow t$ with $t=\ln[(\sqrt {1+v^{2}}+1)/2]$.
This reduces Eq.\ (\ref{D64}) to
\begin{equation}
\Delta=\int_{0}^{\infty}du {1 \over u \sqrt {1+u^{2}}}
\ln[{\sqrt{ 1+v^{2}}+1 \over 2}] .
\label{D65}
\end{equation}
By changing variables via $u \rightarrow t$ with $t=\ln[(\sqrt {1+u^{2}}+1)/2]$,
we eventually obtain $\Delta$ in a form involving a well known integral,
\begin{equation}
\Delta={1 \over 2} \int_{0}^{\infty}dt  {t \over e^{t}-1}=
{\pi^{2} \over 12} .
\label{D66}
\end{equation}
By Eqs.\ (\ref{D66}), (\ref{D63}), and (\ref{D46}), we obtain
\begin{equation}
A_{x}^{(2)}={\pi^{2} \over 12}+{\ln^{2}(2) \over 2} . 
\label{D67}
\end{equation}
Equations (\ref{A_xn}), (\ref{D46}), (\ref{D56}), and (\ref{D67}) yield
our final result for the constant $A_{x}$ anticipated in Eq.\ (\ref{A_x1}).

Next, we compute the integral $A_{z}^{(2)}$ in Eq.\ (\ref{D44}).
For this purpose, consider the integral
\begin{eqnarray}
A_{z}^{(2)}(t) & = &
\int_{0}^{t}du {1-\Phi(u) \over u \sqrt{ 1+u^{2}}}\ln(u \sqrt{ 1+u^{2}})
\nonumber \\
& & +\int_{t}^{\infty}du {1 \over u \sqrt {1+u^{2}}} \ln(u \sqrt {1+u^{2}}) ,
\label{D68}
\end{eqnarray}
For $t=1$, $A_{z}^{(2)}(t=1)=A_{z}^{(2)}$ is the desired integral. It is easy to
show, along the lines we used to derive Eqs.\ (\ref{D59}) and (\ref{D60}), that
\begin{equation}
A_{z}^{(2)}=A_{z}^{(2)}(t)+{[\ln(t \sqrt{ 1+t^{2}})]^{2} \over 2}
-{[\ln {\sqrt 2}]^{2}\over 2}, 
\label{D69}
\end{equation}
for {\it any} $t>0$. Thus, by Eqs.\ (\ref{D68}) and (\ref{D69}), 
with $t=\epsilon \rightarrow 0$,
\begin{eqnarray}
A_{z}^{(2)}&=&\lim_{\epsilon \rightarrow 0}
\bigg[
\int_{\epsilon}^{\infty}du {1 \over u \sqrt {1+u^{2}}}
\ln(u \sqrt {1+u^{2}})  +{\ln^{2}( \epsilon)\over 2}
 \bigg] \nonumber \\
& &  -  {\ln^{2}(2)  \over 8} .
\label{D70}
\end{eqnarray}
Next, by Eqs.\ (\ref{D70}) and (\ref{D61}),
\begin{eqnarray}
& & A_{z}^{(2)}-A_{x}^{(2)}= \nonumber \\
& & \int_{0}^{\infty}du {1 \over u \sqrt {1+u^{2}}}\ln( \sqrt{ 1+u^{2}})        
 - {\ln^{2}(2) \over 8} . 
\label{D71}
\end{eqnarray}
By the change of variables $u \rightarrow t$, with $u=\sqrt{e^{2t}-1}$,
we find $A_{z}^{(2)}-A_{x}^{(2)}=S_{1}+S_{2}-\ln^{2}(2) / 8$,
where $S_{1}$ and $S_{2}$ are two well known integrals,
$S_{1}=\int_{0}^{\infty}dt~t/(e^{t}+1)=\pi^{2}/12$, and
$S_{2}=\int_{0}^{\infty}dt~t/(e^{2t}-1)=
{1\over 4}\int_{0}^{\infty}dx x/(e^{x}-1)={1\over 4}(\pi^{2}/6)$.
Thus,
\begin{equation}
A_{z}^{(2)}-A_{x}^{(2)}={\pi^{2} \over 8}-{\ln^{2}(2) \over 8}. 
\label{D72}
\end{equation}
By Eqs.\ (\ref{D72}) and (\ref{D67}),
\begin{equation}
A_{z}^{(2)}={5 \pi^{2} \over 24}+{3\ln^{2}(2) \over 8} . 
\label{D73}
\end{equation}
Equations (\ref{D42}), (\ref{D46}), (\ref{D56}), and (\ref{D73})  yield
our final result for the constant $A_{z}$ anticipated in Eq.\ (\ref{A_z1}).

We now proceed to discuss calculations yielding the 
values of the numerical constants $B_{x}$ and $B_{z}$
quoted in Eqs.\ (\ref{B_x1}) and (\ref{B_z1}).  For $B_{x}$, by
(\ref{B_x}) and (\ref{Fdef}), we obtain
\begin{equation}
B_{x}=\int_{0}^{\pi}dw
\bigg[{\big(A_{x}^{(3)}(t)\big)_{t=1/w \sqrt {p(w)}} \over w \sqrt{p(w)}}
-{\big(A_{x}^{(3)}(t)\big)_{t=1/w} \over w}\bigg] .
\label{D74}
\end{equation}
By using (\ref{D60}), i.e., 
$A_{x}^{(3)}(t)= A_{x}^{(3)}-\ln(t)=\ln(2)-\ln(t)$, we find
\begin{eqnarray}
B_{x}& = &\bigg[\int_{\epsilon}^{\pi}dw
{\ln(2)+\ln(w \sqrt{p(w)}) \over w \sqrt{p(w)}} \nonumber \\
& & -\int_{\epsilon}^{\pi}dw {\ln(2)+\ln(w) \over w}
\bigg]_{\epsilon \rightarrow 0} . 
\label{D75}
\end{eqnarray}
Next, in the first integral above, we change from $w$ to 
$z=w \sqrt{p(w)}=2 \sin(w/2)$,
whereas in the second integral we change the variable $w$ into $z$. 
After rearranging the expression thus obtained, we find
\begin{eqnarray}
B_{x}&=& \int_{0}^{2}dz \bigg[{1\over \sqrt{1-(z/2)^{2}}}-1 \bigg]
{\ln(2)+\ln(z) \over z} \nonumber \\
& & -  \int_{2}^{\pi}dz{\ln(2)+\ln(z) \over z} . 
\label{D76}
\end{eqnarray}
For convenience, we change variables via $z \rightarrow x=z/2$ and thus obtain
\begin{eqnarray}
B_{x}& = & B^{(1)}+2 \ln(2) B^{(2)}
-{\ln^{2}(\pi) \over 2}\nonumber \\
& & -[\ln(2)][\ln(\pi)]+{3\ln^{2}(2) \over 2} ,
\label{D77}
\end{eqnarray}
with 
\begin{equation}
B^{(1)}=\int_{0}^{1}dx \bigg[{1\over \sqrt{1-x^{2}}}-1 \bigg]\ln(x), 
\label{D78}
\end{equation}
and
\begin{equation}
B^{(2)}=\int_{0}^{1}dx \bigg[{1\over \sqrt{1-x^{2}}}-1 \bigg]. 
\label{D79}
\end{equation}
A similar formula can be derived for the constant 
$B_{z}=\overline B_{z}+B_{x}$ introduced in Sec.\ \ref{sec:zlimit}
of this Appendix.
By Eqs.\ (\ref{F(v,t)}) and (\ref{D30})), we find for $\overline B_{z}$
\begin{equation}
\overline B_{z}  = 
\bigg[\int_{\epsilon}^{\pi}du
{\gamma+\ln(u \sqrt{p(u)}) \over u \sqrt{p(u)}} 
 -\int_{\epsilon}^{\pi}du {\gamma+\ln(u) \over u}
\bigg]_{\epsilon \rightarrow 0} . 
\label{D80}
\end{equation}
By treating Eq.\ (\ref{D80}) in exactly the same way we treated above
Eq.\ (\ref{D75}), we eventually find
\begin{eqnarray}
\overline B_{z}& = &
B^{(1)}+[\ln(2)+\gamma]B^{(2)}+
{\ln^{2}(2) \over 2} \nonumber \\
& & +\gamma \ln(2)-
{\ln^{2}(\pi) \over 2}-\gamma\ln(\pi) . 
\label{D81}
\end{eqnarray}
To complete our calculation, we need the integrals
$B^{(1)}$ and $B^{(2)}$ in Eqs.\ (\ref{D78}) and (\ref{D79}).
$B^{(2)}$, Eq.\ (\ref{D79}), can be calculated by elementary 
integration methods yielding
\begin{equation}
B^{(2)}=\ln(2) .
\label{D82}
\end{equation}
On the other hand, the integral $B^{(1)}$, Eq.\ (\ref{D78}), is not elementary. 
As detailed below, we find
\begin{equation}
B^{(1)}=-{\pi^{2} \over 24}+{\ln^{2}(2) \over 2} . 
\label{D83}
\end{equation}
The values of  $B_{x}$ and $B_{z}=\overline B_{z}+B_{x}$
quoted in Eqs.\ (\ref{B_x1}) and (\ref{B_z1}), directly follow
from Eqs.\ (\ref{D77}), (\ref{D81}), (\ref{D82}), and (\ref{D83}).
It remains to outline the calculation yielding the 
value of $B^{(1)}$ in Eq.\ (\ref{D83}).
It is obtained by relating the integral $B^{(1)}$, Eq.\ (\ref{D78}), 
to the integral
$A_{x}^{(2)}$, Eq.\ (\ref{Ax22}), computed in Eq.\ (\ref{D67}).
We find that
\begin{equation}
B^{(1)}=-{\pi^{2} \over 8}+A_{x}^{(2)} ,
\label{D84}
\end{equation}
To derive relation (\ref{D84}), we apply the Cauchy residue theorem to 
the complex function $f(z)=\ln(z)/z \sqrt{1-z^{2}}$ along the
same contour that has been used before to derive Eq.\ (\ref{A_x11})
[see the text following Eq.\ (\ref{A_zgamma})].
This yields the relation
\begin{eqnarray}
&&\int_{\epsilon}^{1}{dx \over x} {1 \over \sqrt{1-x^{2}}}\ln(x)= \nonumber
\\
&& -{\pi^{2} \over 8}+\int_{\epsilon}^{\infty}{du \over u}
{1\over \sqrt{ 1+u^{2}}}\ln(u)
+\phi (\epsilon) ,
\label{D85}
\end{eqnarray}
where $\phi (\epsilon) \rightarrow 0$ as $\epsilon \rightarrow 0$. 
Equation (\ref{D85}) is identical to
\begin{eqnarray}
&&\int_{\epsilon}^{1}{dx \over x} \bigg[{1 \over \sqrt{1-x^{2}}}-1\bigg]\ln(x)
= \nonumber \\
& & -{\pi^{2} \over 8}+
\int_{\epsilon}^{\infty}{du \over u} {1\over \sqrt{ 1+u^{2}}}\ln(u)
+{\ln^{2}(\epsilon) \over 2}
+\phi (\epsilon) .
\label{D86}
\end{eqnarray}
By recalling here Eq.\ (\ref{D61}) and taking the limit $\epsilon \rightarrow 0$
in Eq.\ (\ref{D86}), we eventually obtain the relation 
(\ref{D84}) used to compute $B^{(1)}$.

\section{Calculation of $\lowercase{g^{(3)} ( {\bf r}, na )}$}
\label{app:g3rna}

In this appendix, we will outline the evaluation of Eq.\ (\ref{g3rnev}) for
$g^{(3)}({\bf r}, na)\equiv g^{(3)}(x,z,na)$.  
We begin with the expression, Eq.\ (\ref{g3rna})
for $g^{(3)}({\bf r}, na)$.  We assume the continuum limit $\Lambda_z
\rightarrow \infty$ and integrate over $q_z$ to obtain
\begin{eqnarray}
g^{(3)} ( {\bf r}, na) &=  &2 l_u^2 \int_0^{\Lambda_x x^*} {d t \over t}
\int_0^{\pi} du {1 - \cos (nu) \over t \sqrt{t^2+u^2 p(u)}} \nonumber \\
& & \times\left[1 - \cos (tx/x^*) e^{-(tz/z^*)  \sqrt{t^2 u^2+p(u)}}
\right].
\label{F1}
\end{eqnarray}
This expression can now be used to evaluate various limits.
\subsection{$x\gg x^*$ and $z=0$}
From Eq.\ (\ref{F1}), we have
\begin{equation}
g^{(3)}(x,0,na)=2 l_u^2 \int_0^{\Lambda_x x^*}{d t \over t} S_n(t)[1 -
\cos( tx/x^*)] ,
\label{F2}
\end{equation}
where $S_n(t)$, defined in Eq.\ (\ref{sndef}), is proportional to $1/t$
for $t\gg 1$.  Thus the integral in Eq.\ (\ref{F2}) is convergent at large
$t$, and we can take the continuum limit $\Lambda_x \rightarrow \infty$. 
Thus, we have
\begin{eqnarray}
g^{(3)}(x,0,na)&=& 2 l_u^2 \int_0^\infty {d t \over t} S_n(t) [1 - \cos (t
x /x^*)] \nonumber \\
& =& 2 l_u^2 \int_0^1{dt\over t}[1-\cos(t x/x^*)] S_n(t)
\nonumber\\
&& + 2 \l_u^2 \int_1^{\infty} {d t \over t}[1-\cos(t x/x^*)]S_n(t) \\
&=& 2 l_u^2 [\ln(x/x^*) + \gamma] S_n(0) \nonumber \\
& & + 2 l_u^2 \int_0^1{dt\over t}[1 -
\cos(t x / x^*)] [S_n(t) - S_n (0)] \nonumber \\
& & + 2 l_u^2 \int_1^{\infty} {dt \over t} [ 1 - \cos (t x / x^*)] S_n(t)
\label{F4}
\end{eqnarray}
The contributions from the $\cos(t x/x^*)$ terms in the 
integrals in Eq.\ (\ref{F4}) vanish when $x/x^* \gg 1$, leaving
\begin{equation}
g^{(3)}(x,0,na) = 2 l_u^2 S_n(0) \ln (D_n x/x^*) ,
\label{F5}
\end{equation}
where
\begin{equation}
D_n = e^{\gamma +  c_n^{(1)} + c_n^{(2)}(\Lambda_x = \infty)}
\label{F6}
\end{equation}
with $c_n^{(1)}$ and $c_n^{(2)}(\Lambda_x)$ given, respectively, by
Eqs.\ (\ref{C5}) and (\ref{C6}).
\subsection{$x=0$, $z\gg z^*$}
Again, we will simplify our discussion by taking the limit $\Lambda_x
\rightarrow \infty$ and integrating over $q_x$ in Eq.\ (\ref{g3rnev}) to
obtain
\begin{eqnarray}
g^{(3)}(0,z,na)= 2 l_u^2 \int_0^{\infty} dt \int_0^\pi du
{1 - \cos( nu) \over t \sqrt{t^2 + u^2 p(u)}} \nonumber \\
\times \left[ 1 - e^{- (t
z/z^*)\sqrt{t^2+u^2 p(u)}} \right].
\label{F7}
\end{eqnarray}
Changing variables via $t = s\sqrt{u^2 p(u)}$, we obtain
\begin{equation}
g^{(3)}(0,z,na)=2 l_u^2 \int_0^{\pi}{1 - \cos (nu)\over \sqrt{u^2 p(u)}}
J[(z/z^*)u^2 p(u)] ,
\end{equation}
where $J(v)$ is defined in Eq.\ (\ref{J(v)}).  Next, using $J(v) = \ln (2
e^{\gamma}) + {\tilde J}(v)$ [Eq.\ (\ref{jvresult})], we obtain
\begin{eqnarray}
g^{(3)}(0,z,na) & = & 2 l_u^2 [\ln(z/z^*) + \gamma +\ln 2]\int_0^{\pi} du
{1 - \cos (nu)\over\sqrt{u^2 p(u)}} \nonumber \\
& & + 2 l_u^2 \int_0^{\pi} du  {1 - \cos (nu)\over\sqrt{u^2 p(u)}} \ln[u^2 p(u)]
\nonumber \\
& &  + 2 l_u^2 \int_0^{\pi} du  {1 - \cos (nu)\over\sqrt{u^2 p(u)}}
{\tilde J}[(z/z^*)u^2 p(u)]
\label{F8}
\end{eqnarray}
In the limit $z/z^* \rightarrow \infty$, ${\tilde J}[(z/z^*)u^2 p(u)]$
tends to zero for all $u$, and we obtain
\begin{equation}
g^{(3)}(0,z,na)= 2 l_u^2 S_n(0) \ln ( E_n z/z^*) ,
\label{F9}
\end{equation}
where
\begin{equation}
E_u = 2 e^{\gamma} \exp\left[ {1 \over S_n(0)} \int_0^{\pi}du{1 - \cos (nu)
\over \sqrt{u^2 p(u)}}\ln [u^2 p(u)]\right]
\label{Du}
\end{equation}

\section{Inter-plane Correlations when $\lowercase{V^u \neq 0}$}
\label{app:inpl-corr}

In this appendix, we will derive Eq.\ (\ref{Sq}) from the expression,
Eq.\ (\ref{Srna}) for $S({\bf r} , na )$.  We begin by looking at the linear
term in $V^u$ in $S({\bf r} , a )$:
\begin{eqnarray}
& & S({\bf r}, a) = {V^u/2 T} \int d^2r_1 
\langle e^{-i q_0 u_z^0 (0)} \nonumber \\
& & \times \left(e^{iq_0[u_z^0 (
{\bf r}_1 )-u_z^1({\bf r}_1 )]} e^{i q_0 a \partial_z u_y^1 ({\bf r}_1 )} 
+ {\rm c.c.} \right)
e^{iq_0 u_z^1 ( {\bf r} )} \rangle_{SC} \\
&& \qquad ={V^u\over 2T}\int d^2 r_1 \langle e^{iq_0 [u_z^1 ( {\bf r} ) - u_z^1 (
{\bf r}_1)]} \nonumber \\
&& \qquad \qquad\times e^{iq_0[u_z^0 ( {\bf r}_1 ) - u_z^0 ( 0 )]} 
e^{i q_0 a\partial_z u_y^1 ( {\bf r}_1 )} \rangle_{SC} \nonumber\\
& & \qquad+ {V^u\over 2T}\int d^2 r_1 \langle e^{iq_0 [u_z^1 ( {\bf r} ) + u_z^1 (
{\bf r}_1)]} \nonumber \\
& & \qquad \qquad \times e^{iq_0[u_z^0 ( {\bf r}_1 ) + u_z^0 ( 0 )]}
\times  e^{i q_0 a \partial_z u_y^1 ( {\bf r}_1 ) }\rangle_{SC} ,
\label{Srn1}
\end{eqnarray}
where the subscript $SC$ indicates that the averages are to be evaluated
with respect to the sliding columnar Hamiltonian, ${\cal H}^{SC}$
of Eq.\ (\ref{HSC}).
Since ${\cal H}^{SC}$ is quadratic in $u_z^n ( {\bf r} )$ 
and $u_y^n ( {\bf r} )$,
the averages in this equation can be performed exactly, and $S_2 ( {\bf r} ,
a )$ can be expressed as an exponential of correlations functions of
$u_z^n ( {\bf r} )$ and $\partial_z u_y^n ( {\bf r} )$.  The second term in 
Eq.\ (\ref{Srn1}) has terms in the exponential proportional to $-q_0^2
\langle [u_z^n ({\bf r} )]^2 \rangle$ for $n=0,1$, which diverge in
thermodynamic limit and cause the exponential to vanish.  Thus only the
first term of Eq.\ (\ref{Srn1}) survives, and we have
\begin{equation}
\label{Sr11}
S({\bf r} , a ) = {V^u\over 2 T} \int d^2 r_1 e^{ - \Phi ({\bf r}_1, {\bf r} )} ,
\end{equation}
where
\begin{eqnarray}
\label{Sr12}
& &\Phi({\bf r}_1,{\bf r})  = \case{1}{2}q_0^2
\langle ({\tilde u}_z^1 ( {\bf r}, {\bf r}_1 )+
{\tilde u}_z^0 ({\bf r}_1 ,0 ) + a \partial_z u_y^1 ( {\bf r}_1 )^2 \rangle \nonumber\\
&& = g^{(2)}({\bf r} - {\bf r}_1 , 0) + g^{(2)} ( {\bf r}_1 , 0 ) \nonumber\\
&& + g^{(2)} ({\bf r},a) +g^{(2)}(0,a) - g^{(2)}({\bf r}- {\bf r}_1,a) -
g^{(2)}({\bf r}_1,a) \nonumber\\
& & + \case{1}{2} a^2 \langle[\partial_z u_y^1 ( {\bf r}_1 )]^2\rangle
\nonumber\\
&& +
a\langle \partial_z u_y^1 ( {\bf r}_1 ) [{\tilde u}_z^1({\bf r}, {\bf r}_1) - 
{\tilde u}_z^0 ( {\bf r}_1, 0 ) ]\rangle ;
\end{eqnarray}
here, $g^{(2)} ({\bf r}, na)$ is defined in Eq.\ (\ref{g2rn2}), and
\begin{equation}
{\tilde u}_z^n ( {\bf r}, {\bf r}' ) = u_z^n ( {\bf r} ) - u_z^n ( {\bf r}' ) .
\end{equation}
This expression is quite complex.  It, however, simplifies considerably
if we set $K_y = 0$ and $B_{uh} = 0$.  Then $g^{(2)} ( {\bf r} , a ) = 0$,
and all cross terms in $u_z^n ( {\bf r} )$ and $u_y^m ( {\bf r}' )$ vanish.  In
this case,
\begin{equation}
\Phi ( {\bf r}_1 , {\bf r} ) = g^{(2)} ( {\bf r} - {\bf r}_1 ) + g^{(2)} ( {\bf r}_1 ) + W_y
,
\end{equation}
where $g^{(2)} ({\bf r} ) = g^{(2)}({\bf r}, 0 )$ [See Eq.\ (\ref{g2rn2})] and
$W_y = q_0^2 a^2 \langle (\partial_z u_y^n )^2 \rangle/2$.

The generalization of Eqs.\ (\ref{Sr11}) and (\ref{Sr12}) to $n>1$ is
straightforward.  The leading contribution to $S({\bf r}, na )$ is
\begin{equation}
S({\bf r},na)= \left( {V^u \over 2T}\right)^n \int d^2 r_1 ... d^2 r_n
e^{- \Phi ( {\bf r}_1, ... , {\bf r}_n , {\bf r} )} ,
\end{equation}
where
\begin{equation} 
\Phi = \case{1}{2} \left\langle \left(\sum_{m=0}^n
{\tilde u}_z^m({\bf r}_m,{\bf r}_{m+1} ) + a \sum_{m=1}^n \partial_z u_y^m ( {\bf r}_m )
\right)^2 \right\rangle_{SC} ,
\end{equation}
where ${\bf r}_0 =0$ and ${\bf r}_{n+1} = {\bf r}$.
This function can be expressed in terms of the reduced correlation
functions $g^{(2)}({\bf r} , n)$ for $u_z^n ( {\bf r} )$ and correlation
functions involving $\partial_zu_y^n ( {\bf r} )$.  In general, it will have
terms connecting all pairs of layers, and the evaluation of $S({\bf r} , na )$
is highly nontrivial.  

We can obtain a useful approximation by setting $K_y=0$ to eliminate all
$g^{(2)}({\bf r} , na )$ for $n \neq 0$ and by setting $B_{uh}=0$ to
eliminate couplings between $u_z^n$ and $u_y^m$.  Even in this
approximation, couplings between distant layers arise from $\langle
\partial_z u_y^n ( {\bf r} ) \partial_z u_y^m ( {\bf r}') \rangle$.  This
function, however, dies off rapidly with $n-m$, and we will set it
equal to zero when $n\neq m$.  [In another publication\cite{OGLT}, we
will use a variational procedure to calculate $S({\bf r} , na)$.  The form of
$S({\bf r} , na )$ evaluated with this procedure is very similar to that
obtained using the above approximations.]  With these approximations, 
\begin{equation}
\Phi({\bf r}_1, ... , {\bf r}_n, {\bf r} ) = q_0^2 \sum_{m=0}^n g^{(2)}({\bf r}_m -
{\bf r}_{m+1} ) + n W_y 
\end{equation}
and
\begin{eqnarray}
S ({\bf r} , na ) &=&{{\tilde V}^u\over 2T}^n \int d^2 r_1 ... d^2 r_n S_2 ( {\bf r}_1
) \nonumber \\
& & \qquad\times S_2 ( {\bf r}_1 - {\bf r}_2 ) ... S_2 ( {\bf r}_n - {\bf r} ) ,
\end{eqnarray}
where
\begin{equation}
S_2 ( {\bf r} ) = \langle e^{i q_0 [ u_z^n ( {\bf r} ) - u_z^n ( 0 ) ]}\rangle
\end{equation}
is the inplane density correlation function and where
\begin{equation}
{\tilde V}^u = V^u e^{- W_y} .
\end{equation}
Fourier transforming this equation, we obtain Eq.\ (\ref{Sq}) in the
text. 

\section{Interaction Energy between Edge Dislocations}
\label{app:interaction}

In this appendix, we evaluate the interaction energy 
$E({\bf r}, na)$ between two dislocations with separation ${\bf x}=({\bf r},na)$
in the limits $x \gg x^*$, $z=0$ and $z \gg z^*$, $x=0$.  To obtain the 
$x \gg x^*$, $z=0$ limit, we must evaluate
\begin{eqnarray}
E_n(x) & = & -\int^{\Lambda_x x^*}_0 {dt \over t} \int_0^{\pi} du
\sqrt{t^2 + u^2 p(u)} \nonumber \\
& \times & \cos [nu] \left(1 - \cos[t x/x^*]\right).
\end{eqnarray}
We isolate the $\ln x$ divergence by adding and subtracting
$\sqrt{u^2 p(u)}$ under the $u$-integral.  This procedure 
yields 
\begin{eqnarray}
E_n(x) = -J_n \ln\left[ C^x_n(\Lambda_x) |x|/x^*\right]
\end{eqnarray}
for $x \gg x^*$.  In the above expression, 
$C^x_n(\Lambda_x) = e^{\gamma} B_n(\Lambda_x)$ with $B_n(\Lambda_x) =
\Lambda_x x^* e^{ {\overline B}_n(\Lambda_x)}$, 
\begin{equation}
{\overline B}_n(\Lambda_x) = \int_0^{\Lambda_x x^*} {dt \over t} \left[
{J_n(t) \over J_n} - 1\right],  
\end{equation}
and 
\begin{equation}
J_n(t) = \int_0^{\pi} du \sqrt{t^2 + u^2p(u)} \cos nu.
\end{equation}

The $z \gg z^*$, $x=0$ limit is obtained in a similar way from 
\begin{eqnarray}
E_n(z) & = & -\int^{\Lambda_x x^*}_0 {dt \over t} \int_0^{\pi} du
\sqrt{t^2 + u^2 p(u)} \nonumber \\
& \times & \cos[nu] \left(1 - \exp\left[-{z \over z^*}t\sqrt{t^2 + u^2p(u)}
\right]\right).
\end{eqnarray}
We find that 
\begin{eqnarray}
E_n(z) = - J_n \ln\left[ C^z_n(\Lambda_x) |z|/z^*\right]
\end{eqnarray} 
scales logarithmically for $z \gg z^*$, where $C^z_n(\Lambda_x) =
B_n(\Lambda_x) e^{ {\overline C}_n^z(\Lambda_x)}$,
\begin{eqnarray}
{\overline C}^z_n = \int_0^1 dy \left[{1-F_n(y) \over y}\right] 
- \int_1^{\infty} dy {F_n(y) \over y},
\end{eqnarray}
and 
\begin{eqnarray}
F_n(y) = {1 \over J_n} \int_0^{\pi} du \cos(nu) 
\sqrt{u^2 p(u)} \exp\left[-y \sqrt{u^2 p(u)}
\right]. \nonumber
\end{eqnarray}
Note that $C_n^{x,z}(\Lambda_x)$ diverge with $\Lambda_x$, and 
thus $E_n({\bf r})$ does not have a well-defined $\Lambda_x 
\rightarrow \infty$ limit.

\end{document}